\documentclass[]{JHEP3}
\pdfoutput = 1
\usepackage{graphicx}
\usepackage{amsmath,amssymb,amscd,amsfonts,bm}

\newcommand{\ME}{M}

\newcommand{\as}{\alpha_{\mathrm{s}}}
\newcommand{\LA}{\mathrm{A}}
\newcommand{\LB}{\mathrm{B}}
\newcommand{\LI}{\mathrm{I}}
\newcommand{\La}{\mathrm{a}}
\newcommand{\Lb}{\mathrm{b}}
\newcommand{\Lc}{\mathrm{c}}
\newcommand{\Ls}{\mathrm{s}}
\newcommand{\LS}{\mathrm{S}}
\newcommand{\mpone}{{m+1}}

\def\ket#1{\big|{#1}\big\rangle}
\def\bra#1{\big\langle{#1}\big|}
\def\brax#1{\big\langle{#1}}   

\def\sket#1{\big|{#1}\big)}
\def\sbra#1{\big({#1}\big|}
\def\sbrax#1{\big({#1}}        

\def\dualL{\raisebox{-5 pt}{$\scriptstyle D$}\!} 

\newbox\charbox
\newbox\slabox
\def\s#1{{      
        \setbox\charbox=\hbox{$#1$}
        \setbox\slabox=\hbox{$/$}
        \dimen\charbox=\ht\slabox
        \advance\dimen\charbox by -\dp\slabox
        \advance\dimen\charbox by -\ht\charbox
        \advance\dimen\charbox by \dp\charbox
        \divide\dimen\charbox by 2
        \raise-\dimen\charbox\hbox to \wd\charbox{\hss/\hss}
        \llap{$#1$}
}}

\title{Parton showers with quantum interference: leading color, spin averaged}

\author{Zolt\'an Nagy \\
Theory Division,
CERN\\
CH-1211 Geneva 23, Switzerland\\
E-mail: \email{Zoltan.Nagy@cern.ch}
}

\author{Davison E. Soper\\
Institute of Theoretical Science\\
University of Oregon\\
Eugene, OR  97403-5203, USA\\
E-mail: \email{soper@uoregon.edu}
}

\abstract{
We have previously described a mathematical formulation for a parton shower based on the approximation of strongly ordered virtualities of successive parton splittings. Quantum interference, including interference among different color and spin states, is included. In this paper, we add the further approximations of taking only the leading color limit and averaging over spins, as is common in parton shower Monte Carlo event generators. Soft gluon interference effects remain with this approximation. We find that the leading color, spin averaged shower in our formalism is similar to that in other shower formulations. We discuss some of the differences.}

\preprint{
CERN-PH-TH/2007-261\\
11 January 2008
}

\begin{document}


\section{Introduction}
\label{sec:Intro}

In Ref.~\cite{NSshower}, we presented a formalism for a mathematical representation of a parton shower that incorporates interference in both spin and color. In this paper, we analyze this formalism in the approximation that we average over parton spins at each step and keep only the leading contributions in an expansion in powers of $1/N_\Lc^2$, where $N_\Lc = 3$ is the number of colors.\footnote{More precisely, we average over the spins of incoming partons at each step and sum over the spins of the outgoing partons.} Our interest is to elucidate the structure of the full shower formulation of Ref.~\cite{NSshower} by examining what happens when the spin-averaged and leading color approximations are imposed. We also anticipate that the approximate shower may be of use in implementing successively better approximations to the full shower including spin and color.

Our main focus is on the splitting functions that would be used to generate the shower in the spin averaged approximation (which is a customary approximation in current parton shower event generators). In our formalism, there are two sorts of splitting functions. The direct splitting functions correspond to the squared amplitude for a parton $l$ to split into daughter partons that, in our notation, carry labels $l$ and $m+1$, where $m+1$ is the total number of final state partons after the splitting. In this paper, we use the spin dependent splitting functions from Ref.~\cite{NSshower} and simply average over the spins of the mother parton and sum over the spins of the daughter partons. We analyze some of the important properties of these functions. We also need interference splitting functions. These correspond to the interference between the amplitude for a parton $l$ to split into partons with labels $l$ and $m+1$ and the amplitude for another parton $k$ to split into partons with labels $k$ and $m+1$. These functions generate leading singularities when parton $m+1$ is a soft gluon. We improve the specifications of Ref.~\cite{NSshower} for this by defining a useful form for certain weight functions $A_{lk}$ and $A_{kl}$ that were assigned the default values 1/2 in Ref.~\cite{NSshower}. We will see that with the improved form for  $A_{ij}$, the total splitting probabilities acquire useful properties in the soft gluon limit.

We will see that when we make the spin-averaged and leading color approximations, the parton shower formalism of Ref.~\cite{NSshower} amounts to something quite similar to standard parton shower event generators. One significant point in common is that the splitting functions are positive. One difference with some standard event generators is that an angular ordering approximation is not needed because the coherence effects that lead to angular ordering are built into the formalism from the beginning, both for initial state and final state splittings. This coherence feature is a natural consequence of a dipole based shower, as in the final state showers of \textsc{Ariadne} \cite{Ariadne} and the $k_T$ option of \textsc{Pythia} \cite{SjostrandSkands} or the showers \cite{Schumann, Weinzierl} based on the Catani-Seymour dipole splitting formalism \cite{CataniSeymour}. Additionally, our formalism differs from others in its splitting functions and its momentum mappings.

\section{Direct spin-averaged splitting functions}
\label{sec:directSFs}

We begin with the splitting functions that correspond to the amplitude for a parton to split times the complex conjugate amplitude for that same parton to split. We follow the notation of Ref.~\cite{NSshower}. Before the splitting, there are partons that carry the labels $\{\La,\Lb,1,\dots,m\}$, where $\La$ and $\Lb$ are the labels of the initial state partons. The momenta and flavors of these partons are denoted by $\{p,f\}_m = \{p_\La,f_\La;\dots;p_m,f_m\}$. The flavors are $\{{\rm g}, {\rm u}, \bar {\rm u}, {\rm d}, \dots\}$, with the initial state flavors $f_\La$ and $f_\Lb$ denoting the flavors coming out of the hard interaction and thus the opposite of the flavors entering the hard interaction. We let $l$ be the label of the parton that splits. After the splitting, there are $m+1$ final state partons. The momenta and flavors of the partons are $\{\hat p,\hat f\}_{m+1}$. We use the label $l$ for one of the daughter partons and the label $m+1$ for the other daughter parton.\footnote{For a final state $q \to q {\rm g}$ splitting, we use $m+1$ for the label of the gluon. For a final state ${\rm g} \to q\,\bar q$ splitting, we use $m+1$ for the label of the $\bar q$.} The partons that do not split keep their labels. However, they donate some of their momenta to the daughter partons so that the daughter partons can be on shell. Thus $\hat p_i \ne p_i$ in general for a spectator parton. The momenta and flavors after the splitting, $\{\hat p, \hat f\}_{m+1}$, are determined by the momenta and flavors before the splitting, $\{p,f\}_m$ and variables $\{\zeta_{\rm p},\zeta_{\rm f}\}$ that describe the splitting.\footnote{When a gluon splits, $\zeta_{\rm f}$ determines whether the daughters are a $({\rm g},{\rm g})$ pair, a $({\rm u},\bar {\rm u})$ pair, {\it etc}. In Ref.~\cite{NSshower}, we defined the splitting variables $\zeta_{\rm p}$ in a rather abstract way, but one could imagine using for $\zeta_{\rm p}$ the virtuality of the daughter parton pair, a momentum fraction, and an azimuthal angle.} A certain mapping
\begin{equation}
\label{eq:Rldef}
\{\hat p, \hat f\}_{m+1} = R_l(\{p,f\}_m, \{\zeta_{\rm p},\zeta_{\rm f}\})
\end{equation}
defined in Ref.~\cite{NSshower} gives the relation.

\FIGURE{
\centerline{\includegraphics[width = 11 cm]{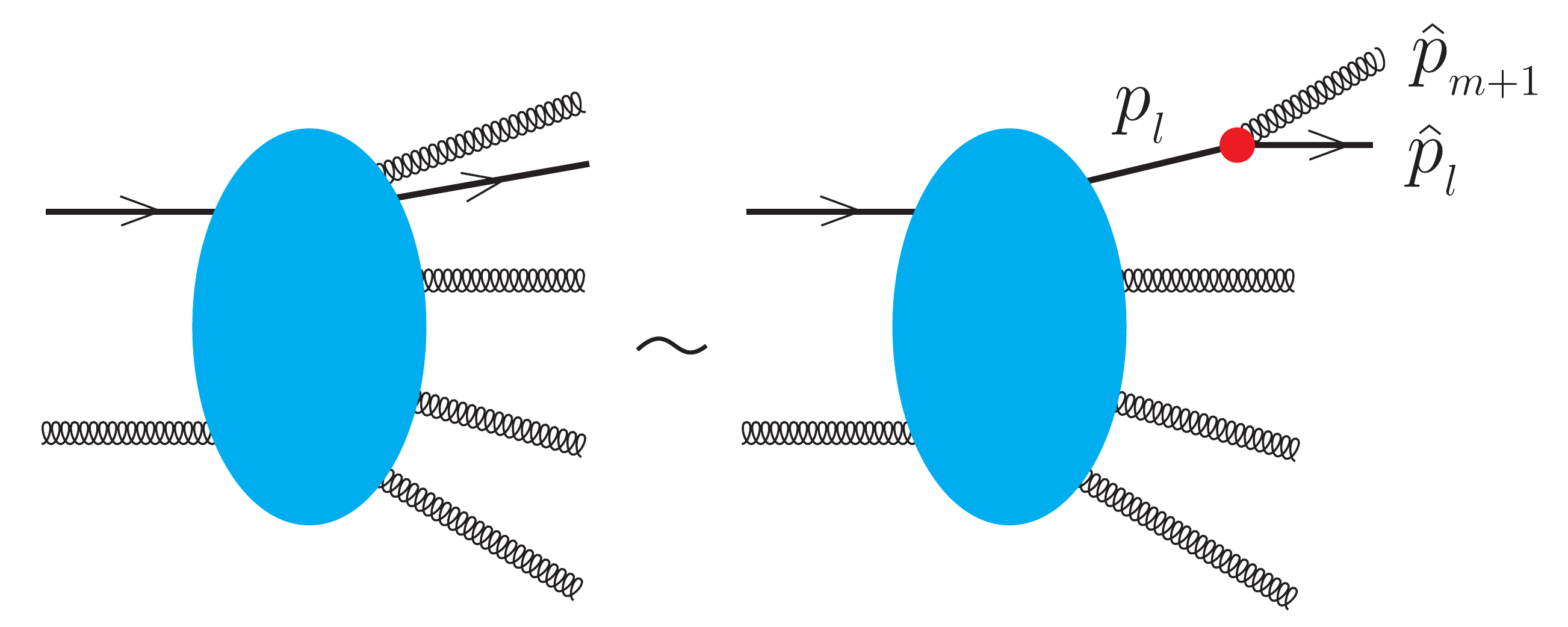}}
\caption{Illustration Eq.~(\ref{eq:splittingamplitudestructure}) for a $qq{\rm g}$ final state splitting. The small filled circle represents the splitting amplitude $v_l$. The mother parton has momentum $\hat p_l + \hat p_{m+1}$, but in the amplitude $\ket{\ME(\{p, f\}_{m})}$, this off-shell momentum is approximated as an on-shell momentum $p_l$.}
\label{fig:qqgFSsplit}
} 

The splitting functions in Ref.~\cite{NSshower} are based on spin dependent splitting amplitudes $v_l$. One starts with the amplitude $\ket{\ME(\{p, f\}_{m})}$ to have $m$ partons. The amplitude is a vector in spin$\otimes$color space. After splitting parton $l$, we have a new amplitude $\ket{\ME_l(\{p, f\}_{m+1})}$ of the form illustrated in Fig.~\ref{fig:qqgFSsplit}
\begin{equation}
\label{eq:splittingamplitudestructure}
\ket{\ME_l(\{\hat p, \hat f\}_{m+1})} = 
t^\dagger_l(f_l \to \hat f_l + \hat f_{m+1})\,
V^\dagger_l(\{\hat p, \hat f\}_{m+1})\, \ket{\ME(\{p, f\}_{m})}
\;\;.
\end{equation}
Here $t^\dagger_l$ is an operator on the color space that simply inserts the daughter partons with the correct color structure. The factor $V^\dagger_l$ is a function the momenta and flavors and is an operator on the spin space. It leaves the spins of the partons other than parton $l$ undisturbed and multiplies by a function $v_l$ that depends on the mother spin and the daughter spins:
\begin{equation}
\label{eq:Vtov}
\bra{\{\hat s\}_{m+1}}
V^\dagger_l(\{\hat p, \hat f\}_{m+1})\ket{\{s\}_m}
= 
\left(\prod_{j\notin\{l,m+1\}} \delta_{\hat s_j,s_j}\right)
v_l(\{\hat p, \hat f\}_{m+1},\hat s_{m+1},\hat s_{l},s_l)
\;\;.
\end{equation}
Thus the splitting is defined by the splitting amplitudes $v_l$, which are derived from the QCD vertices.

We can illustrate this for the case of a final state $q \to q + {\rm g}$ splitting, for which we define
\begin{equation}
\begin{split}
\label{eq:qqgFSamplitude}
v_l(\{\hat p, \hat f\}_{m+1},&\hat s_{m+1},\hat s_{l},s_l) =
\\ &
\sqrt{4\pi\as}\,
 \varepsilon_\mu(\hat p_{m+1},\hat s_{m+1};\hat Q)^*\,
\frac{
\overline U({\hat p_l,\hat s_l})\gamma^\mu 
  [\s{\hat p}_l + \s{\hat p}_{m+1} + m(f_l)] \s{n_l} U({p_l,s_l})}
{[(\hat p_l + \hat p_{m+1})^2 - m^2(f_l)]\,2p_l\!\cdot\! n_l}
\;\;.
\end{split}
\end{equation}
There are spinors for the initial and final quarks. There is a polarization vector for the daughter gluon, defined in timelike axial gauge so that $\hat p_{m+1} \cdot \varepsilon = \hat Q \cdot \varepsilon = 0$. Here $\hat Q$ is the total momentum of the final state partons, which is the same before and after the splitting. There is a vertex $\gamma^\mu$ for the $qq{\rm g}$ interaction. There is a propagator for the off shell quark with momentum $\hat p_l + \hat p_{m+1}$. So far, this is exact. Finally, there is an approximation that applies when the splitting is nearly collinear or soft. We approximate $\hat p_l + \hat p_{m+1}$ by $p_l$ in the hard interaction and insert a projection $\s{n}_l/2 p_l\cdot n_l$ onto the ``good'' components of the Dirac field. This projection uses a lightlike vector $n_l$ that lies in the plane of $\hat Q$ and $p_l$,
\begin{equation}
\label{eq:nldef}
n_l = Q
-\frac{Q^2}
{Q\!\cdot\! p_l
+ \sqrt{(Q\!\cdot\! p_l)^2 - Q^2\, m^2(f_l)}}\
p_l
\;\;.
\end{equation}

\FIGURE{
\centerline{\includegraphics[width = 10 cm]{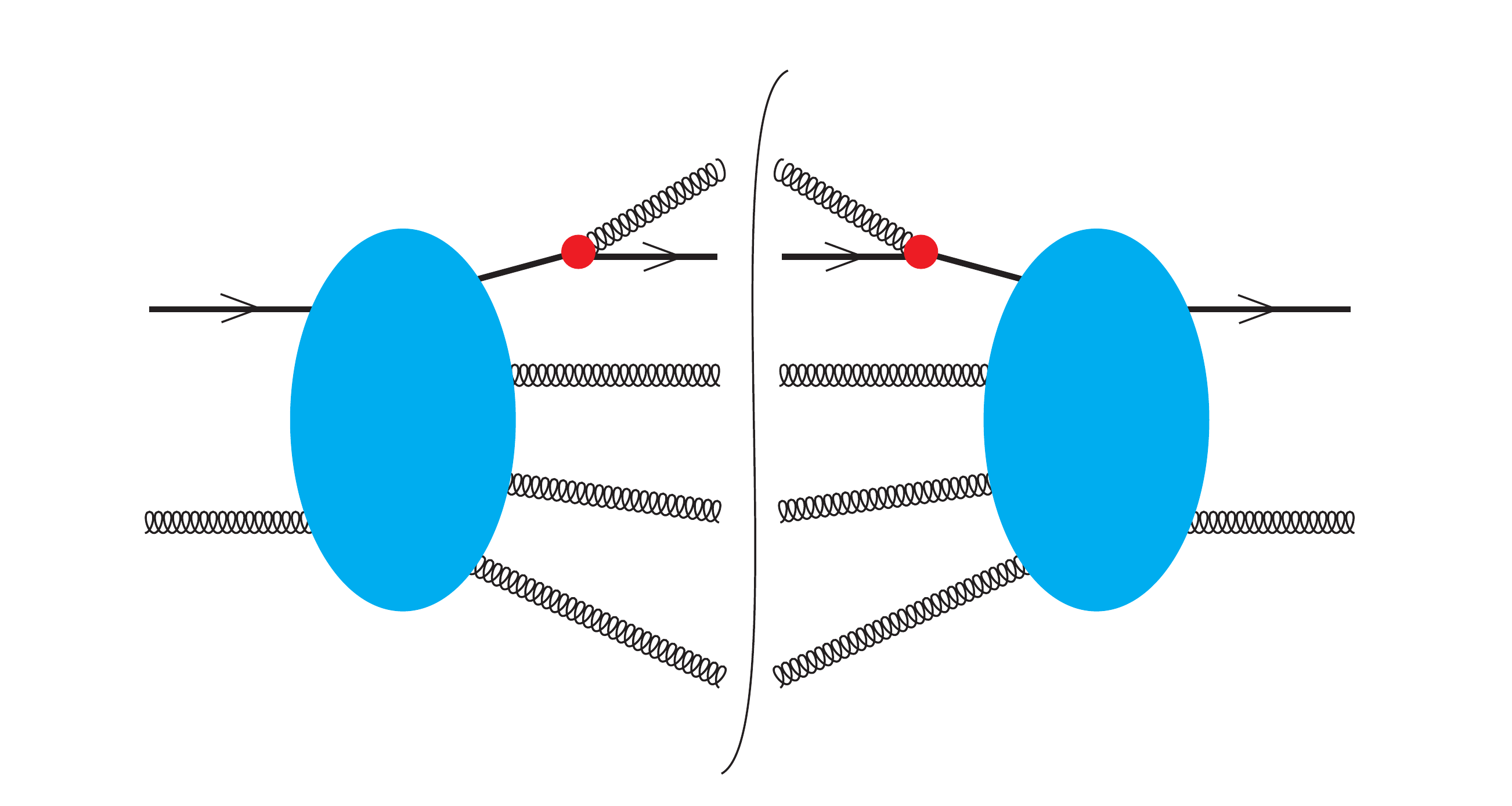}}
\caption{Illustration of how $v_l$ times $v_l^*$ appears in the calculation of the approximate matrix element $\ket{\ME_l(\{\hat p, \hat f\}_{m+1})}$ times its complex conjugate, $\bra{\ME_l(\{\hat p, \hat f\}_{m+1})}$. If we average over spins, we need to multiply $\ket{\ME(\{p, f\}_{m})}$ times $\bra{\ME(\{p, f\}_{m})}$ by $\overline W_{ll}$, Eq.~(\ref{eq:barwlldef}).}
\label{fig:qqgsplitsq}
} 

With one exception, the direct splitting functions in Ref.~\cite{NSshower} are products of a splitting amplitude, $v_l$, times a complex conjugate splitting amplitude, $v_l^*$,
\begin{equation}
v_l(\{\hat p, \hat f\}_{m+1},\hat s_{m+1},\hat s_{l},s_l)\
v_l(\{\hat p, \hat f\}_{m+1},\hat s_{m+1}',\hat s_{l}',s_l')^*
\;\;.
\end{equation}
The calculation of $\ket{\ME_l(\{\hat p, \hat f\}_{m+1})}$ times $\bra{\ME_l(\{\hat p, \hat f\}_{m+1})}$ using $v_l \times v^*_l$ is illustrated in Fig.~\ref{fig:qqgsplitsq}. In this calculation, in general, we have to keep track of two spin indices, $s$ and $s'$ for each parton in order to describe quantum interference in the spin space. However, in this paper we make an approximation. We set $s' = s$ for each parton, sum over the daughter parton spins and average over the mother parton spins. Thus we use a splitting function\footnote{The function $\overline W_{ll}$ here is the same as $\overline w_{ll}$ in Ref.~\cite{NSshower}.}
\begin{displaymath}
\overline W_{ll} = \frac{1}{2}\sum_{\hat s_l,\hat s_{m+1}, s_l}
|v_l(\{\hat p, \hat f\}_{m+1},\hat s_{m+1},\hat s_{l},s_l)|^2
\end{displaymath}
for any flavor combination allowed with our conventions for assigning the labels $l$ and $m+1$ except for a final state ${\rm g} \to {\rm g} + {\rm g}$ splitting, for which we do something slightly different because the two gluons are identical. We make manifest the definition of which flavor combinations are allowed by defining
\begin{equation}
\label{eq:Sldef}
S_l(\{\hat f\}_{m+1}) = \left\{
\begin{array}{cl}
1/2\;, &  l \in \{1,\dots,m\},\ \hat f_l = \hat f_{m+1} = {\rm g}  \\
  1\;, &  l \in \{1,\dots,m\},\ \hat f_l \ne {\rm g}, \hat f_{m+1} = {\rm g}  \\
  0\;, &  l \in \{1,\dots,m\},\ \hat f_l = {\rm g}, \hat f_{m+1} \ne {\rm g}  \\
  1\;, &  l \in \{1,\dots,m\},\ \hat f_l = q, \hat f_{m+1} = \bar q  \\
  0\;, &  l \in \{1,\dots,m\},\ \hat f_l = \bar q, \hat f_{m+1} = q  \\
  1\;, &  l \in \{\La,\Lb\}
\end{array}
\right.
\;\;.
\end{equation}
This is 1 for the allowed combinations, 0 otherwise, with a statistical factor 1/2 for a final state ${\rm g} \to {\rm g} + {\rm g}$ splitting. The complete definition of $\overline W_{ll}$ is then
\begin{equation}
\begin{split}
\label{eq:barwlldef}
\overline W_{ll} ={}&
S_l(\{\hat f\}_{m+1})\
\frac{1}{2}\sum_{\hat s_l,\hat s_{m+1}, s_l}
\bigg\{
|v_l(\{\hat p, \hat f\}_{m+1},\hat s_{m+1},\hat s_{l},s_l)|^2
\\ & + 
\theta(l \in \{1,\dots,m\},\hat f_l = \hat f_m = {\rm g})\
\\&\times
\Big[
|v_{2,l}(\{\hat p, \hat f\}_{m+1},\hat s_{m+1},\hat s_{l},s_l)|^2
-
|v_{3,l}(\{\hat p, \hat f\}_{m+1},\hat s_{m+1},\hat s_{l},s_l)|^2\,
\Big]\bigg\}
\;\;.
\end{split}
\end{equation}
The second term applies for a final state ${\rm g} \to {\rm g} + {\rm g}$ splitting and is arranged to keep the total splitting probability the same but associate the leading soft gluon singularity with gluon $m+1$ rather than gluon $l$. The functions $v_{2,l}$ and $v_{3,l}$ are defined in Sec.~\ref{sec:gggFS}.

The form of the splitting amplitude $v_l$ depends on the type of partons that are involved. However, there is a common result in the limit $\hat p_{m+1} \to 0$ whenever parton $m+1$ is a gluon. In this limit, $v_l$ is given by the eikonal approximation,
\begin{equation}
v_l^{\rm eikonal}(\{\hat p, \hat f\}_{m+1},\hat s_{m+1},\hat s_{l},s_l)
= \sqrt{4\pi\as}\, \delta_{\hat s_l,s_l}\,
\frac{
\varepsilon(\hat p_{m+1}, \hat s_{m+1};Q)^*
\!\cdot\! \hat p_l}
{\hat p_{m+1}\!\cdot\! \hat p_l}
\;\;.
\end{equation}
The soft gluon limit of $\overline W_{ll}$ is then
\begin{equation}
\begin{split}
\label{eq:wlleikonal}
\overline W_{ll}^{\, \rm eikonal} = {}&
4\pi\as\ 
\frac{
 \hat p_l\cdot D(\hat p_{m+1};\hat Q) \cdot  \hat p_l}
{(\hat p_{m+1}\!\cdot\! \hat p_l)^2}\
\;\;.
\end{split}
\end{equation}
Here $D_{\mu\nu}$ is the sum over $\hat s_{m+1}$ of $\varepsilon_\mu \varepsilon_\nu^*$,
\begin{equation}
\label{eq:Dmunu}
D_{\mu\nu}(\hat p_\mpone,\hat Q) =
- g_{\mu\nu} 
+ \frac{\hat p_\mpone^\mu \hat Q^\nu + \hat Q^\mu \hat p_\mpone^\nu}
{\hat p_\mpone\cdot \hat Q}
- \frac{\hat Q^2 \hat p_\mpone^\mu \hat p_\mpone^\nu}{(\hat p_\mpone\cdot \hat Q)^2}
\;\;.
\end{equation}

The function $\overline W_{ll}$ and its approximate form $\overline W_{ll}^{\rm eikonal}$ give the dependence of the splitting operator on momentum and spin for a given set of parton flavors. The partons also carry color. In Ref.~\cite{NSshower} there is a separate factor that gives the color dependence. This factor is an operator on the color space that we can call $t^{\dagger}_l \otimes t^{}_l$, where $t^{\dagger}_l$ is the operator in Eq.~(\ref{eq:splittingamplitudestructure}), which inserts the proper color matrix into the amplitude, and $t^{}_l$ inserts the proper color matrix into the complex conjugate amplitude.\footnote{In Ref.~\cite{NSshower}, we write $t^\dagger_l(f_l \to f_l + {\rm g})$ for the operator that we here call just $t^\dagger_l$ and we denote the operator $t^{\dagger}_l \otimes t^{}_l$ by ${\cal G}(l,l)$.} So far, we do not make any approximations with respect to color. In Sec.~\ref{sec:color}, we will make the approximation of keeping only the leading color conributions.

We now turn to a more detailed discussion of $\overline W_{ll}$ for particular cases.

\subsection{Final state $q \to q + {\rm g}$ splitting}
\label{sec:qqgFS}

Let us look at $\overline W_{ll}$ for a final state $q \to q + {\rm g}$ splitting,
\begin{equation}
\begin{split}
\overline W_{ll} ={}& \frac{4\pi\as}{2(p_l\cdot n_l)^2}\,
\frac{1}{(2\,\hat p_l\cdot \hat p_\mpone)^2}\
D_{\mu\nu}(\hat p_\mpone,\hat Q)
\\&\times
\frac{1}{4}{\rm Tr}\left[
(\s{\hat p}_l + m)\gamma^\mu (\s{\hat p}_l + \s{\hat p}_\mpone + m)\s{n}_l
(\s{p}_l + m)\s{n}_l(\s{\hat p}_l + \s{\hat p}_\mpone + m)\gamma^\nu
\right]
\;\;.
\end{split}
\end{equation}
Here $m = m(f_l)$ is the quark mass, the lightlike vector $n_l$ is given by Eq.~(\ref{eq:nldef}), and $D_{\mu\nu}$ is given by Eq.~(\ref{eq:Dmunu}). It will be convenient to examine the dimensionless function
\begin{equation}
\label{eq:Fdef}
F \equiv \frac{\hat p_l\cdot \hat p_\mpone}{4\pi\as}\
\overline W_{ll}
\;\;.
\end{equation}
The limiting behavior of $F$ as the gluon $m+1$ becomes soft, $\hat p_{m+1} \to 0$, is simple. Then the eikonal approximation applies and we obtain from Eq.~(\ref{eq:wlleikonal})
\begin{equation}
\label{eq:Feikonal}
F_{\rm eikonal} = 
\frac{\hat p_l \cdot D(\hat p_\mpone,\hat Q) \cdot \hat p_l}
{\hat p_l\cdot \hat p_\mpone}\
\;\;.
\end{equation}
The full behavior of $F$ is more complicated,
\begin{equation}
\begin{split}
\label{eq:qqgFSsubtracted}
F ={}& 
[1 + h(y,a_l,b_l)]\,F_{\rm eikonal} 
+ \frac{\hat p_\mpone\cdot n_l}{p_l\cdot n_l}
\;\;.
\end{split}
\end{equation}
Here 
\begin{equation}
\begin{split}
h(y,a_l,b_l) ={}& \frac{1 + y + \lambda r_l}{1 + r_l}
+ \frac{2 a_l y}{\lambda r_l ( 1 + r_l)}
-1
\;\;,
\end{split}
\end{equation}
where
\begin{equation}
\begin{split}
\label{eq:hdefs}
y = {}& \frac{2 \hat p_l\cdot \hat p_\mpone}{2p_l\cdot \hat Q}
\;\;,
\\
a_l ={}& \frac{\hat Q^2}{2p_l\cdot \hat Q}
\;\;,
\\
b_l ={}& \frac{m^2}{2p_l\cdot \hat Q}
\;\;,
\\
r_l ={}& \sqrt{1 - 4 a_l b_l}
\;\;,
\\
\lambda = {}&
\frac{\sqrt{(1+y)^2 - 4 a_l (y + b_l)}}{r_l}
\;\;.
\end{split}
\end{equation}

The eikonal approximation to $F$ will turn out to be significant in our analysis when we incorporate the effect of soft-gluon interference graphs. We will find that it is of some importance for the numerical good behavior of the splitting functions including interference that
\begin{equation}
\label{eq:qqgFSpositivity}
F -  F_{\rm eikonal} \ge 0
\;\;.
\end{equation}
To see that this property holds we note first that  ${\hat p_\mpone\cdot n_l}/{p_l\cdot n_l}$ is non-negative. Remarkably, $h(y,a_l,b_l) \ge 0$ also. To prove this, we first note that
\begin{equation}
h(y,a_l,0) = \frac{(\lambda - 1 + y)^2 + 4 y}{4\lambda}\ 
\;\;,
\end{equation}
so that $h(y,a_l,0) \ge 0$. Then we show that $h(y,a_l,b_l) - h(y,a_l,0) \ge 0$ by simply making plots of this function. This establishes the positivity property Eq.~(\ref{eq:qqgFSpositivity}).

We now examine $F$ further under the assumption that $m=0$. We write $F$ as a function of the dimensionless virtuality variable $y$, and a momentum fraction\footnote{Note that there are many different ways to define a momentum fraction variable. The value of the splitting function for a given choice of daughter parton momenta does not depend on the momentum fraction variable that one uses to label these momenta. We have taken a simple definition of $z$ in order to display results in a graph.}
\begin{equation}
\label{eq:zFSdef}
z = \frac{\hat p_\mpone \cdot n_l}{(\hat p_\mpone + \hat p_l)\cdot n_l}
\;\;.
\end{equation}
It is also convenient to use an auxiliary momentum fraction variable
\begin{equation}
\label{eq:xdefFS}
x = \frac{\hat p_\mpone\cdot \hat Q}{(\hat p_\mpone + \hat p_l)\cdot \hat Q}
= \frac{\lambda}{1+y}\ z + 
\frac{2 a_l y}{(1+y)(1+y+\lambda)}
\;\;,
\end{equation}
where, for $m=0$, $\lambda = \sqrt{(1+y)^2  - 4 a_l y}$. Using these variables, 
\begin{equation}
\label{eq:FeikFS}
F_{\rm eikonal} = 2\ \frac{1-x}{x}
-\frac{2 a_l y}{x^2 (1+y)^2}
\;\;,
\end{equation}
and
\begin{equation}
\begin{split}
\label{eq:qqgFSmzero}
F ={}& 
\left[1 + \frac{(\lambda - 1 + y)^2 + 4 y}{4\lambda}\right]
F_{\rm eikonal} 
+ \frac{1}{2}\ z [1 + y + \lambda]
\;\;.
\end{split}
\end{equation}
As $y \to 0$, $F$ must turn into the Altarelli-Parisi function for this splitting, 
\begin{equation}
F_{\rm AP}(z) = \frac{1 + (1-z)^2}{z}
\;\;.
\end{equation}
Indeed, the derivation given above is one way to derive the Altarelli-Parisi function. We illustrate how $F(z,y,a_l,b_l)$ at $b_l = 0$ approaches $F_{\rm AP}(z)$ in Fig.~\ref{fig:qqgFSplot}.

\FIGURE{
\centerline{\includegraphics[width = 10 cm]{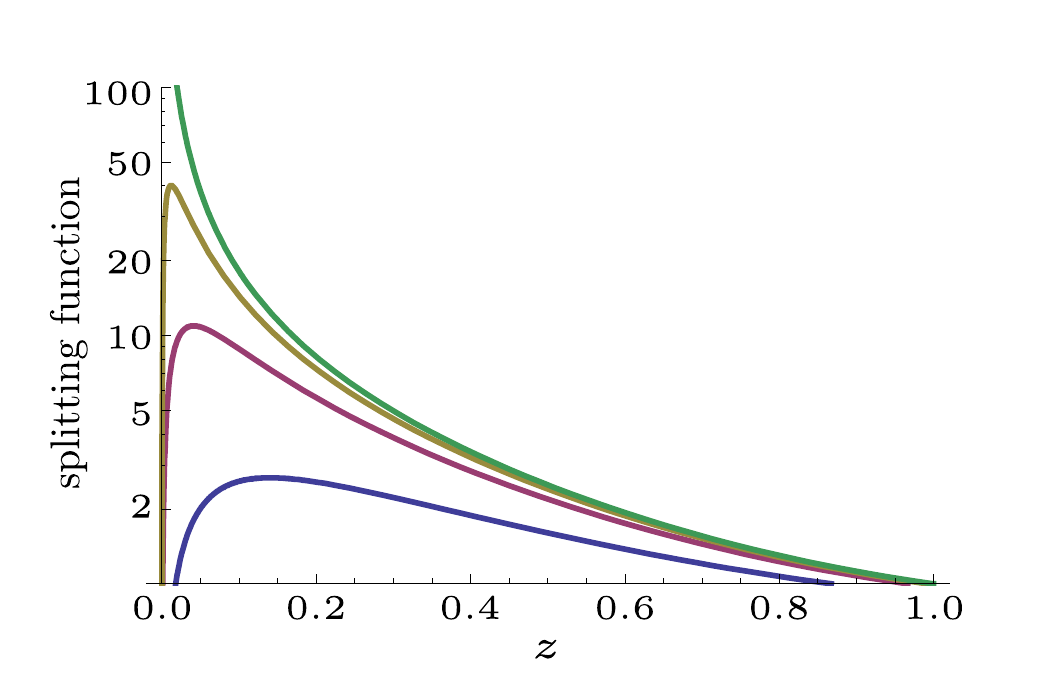}}
\caption{The spin averaged splitting function $F$ defined in Eq.~(\ref{eq:qqgFSmzero}) for a final state $q \to q + {\rm g}$ splitting, plotted versus the momentum fraction $z$ of the gluon, as defined in Eq.~(\ref{eq:zFSdef}). The quark is taken to be massless and we set $a_l = 4$. The four curves are for, from bottom to top, $y = 0.03, 0.01, 0.001, 0$. For $y = 0$, the result is the Altarelli-Parisi function, $[1 + (1-z)^2]/z$.}
\label{fig:qqgFSplot}
} 

\subsection{Initial state $q \to q + {\rm g}$ splitting}
\label{sec:qqgIS}

Here we consider an initial state $q \to q + \mathrm g$ splitting, as illustrated in Fig.~\ref{fig:qqgISsplit}. For notational convenience, we let it be parton ``a'' that splits, so $l = \La$. We allow both parton ``a'' and parton ``b'' to have masses, $m_\La \equiv m(f_\La)$ and $m_\Lb \equiv m(f_\Lb)$. One could, of course, choose these masses to be zero. Parton $m+1$ is a (massless) gluon. The shower evolution for initial state particles runs backwards in physical time. Parton ``a'', which carries momentum $p_\La$ into the hard interaction, splits into the final state gluon with momentum $p_\mpone$ and an initial state parton that carries momentum $\hat p_\La$ into the splitting. For a nearly collinear splitting, $p_\La \approx \hat p_\La - \hat p_\mpone$. In physical time, it is the initial state parton with momentum $\hat p_\La$ that splits. 

\FIGURE{
\centerline{\includegraphics[width = 7.5 cm]{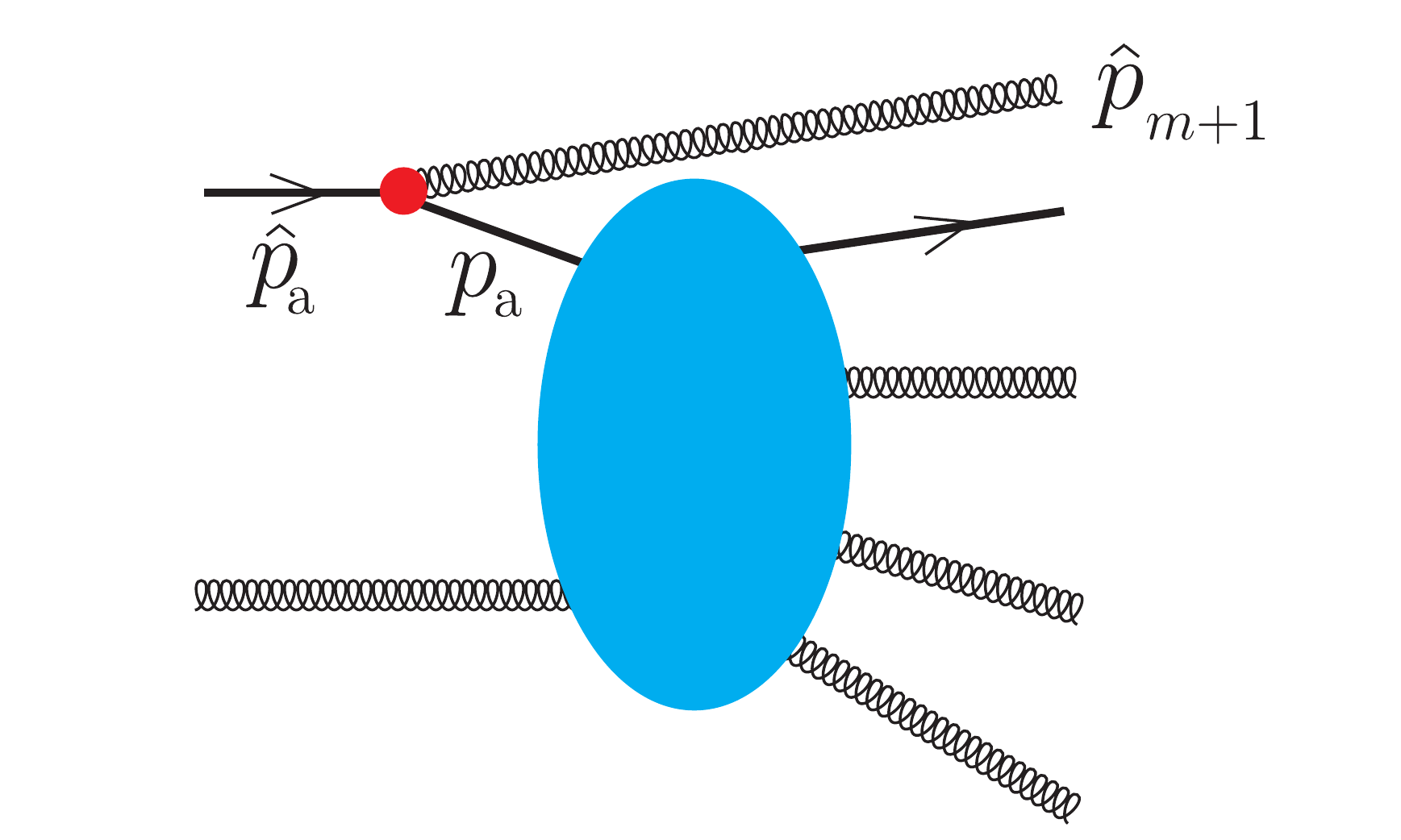}}
\caption{An initial state $q \to q + {\rm g}$ splitting.}
\label{fig:qqgISsplit}
} 

Following Ref.~\cite{NSshower}, we define the kinematics using lightlike vectors $p_\LA$ and $p_\LB$ that are lightlike approximations to the momenta of hadrons A and B, respectively, with $2p_\LA \cdot p_\LB = s$. The momenta of the partons that enter the hard scattering, $p_\La$ and $p_\Lb$, are defined using momentum fractions $\eta_\La$ and $\eta_\Lb$. After the splitting, the momentum fractions are $\hat \eta_\La$ and $\hat \eta_\Lb$. Because parton ``a'' splits, $\hat \eta_\La \ne \eta_\La$. However, with our kinematics, the momentum fraction of parton ``b'' remains unchanged: $\hat \eta_\Lb = \eta_\Lb$. The initial state parton momenta are defined to be
\begin{equation}
\begin{split}
p_\La ={}& \eta_\La p_\LA + \frac{m^2_\La}{\eta_\La s}\, p_\LB
\;\;,
\\
p_\Lb ={}& \eta_\Lb p_\LB + \frac{m^2_\Lb}{\eta_\Lb s}\, p_\LA
\;\;,
\\
\hat p_\La ={}& \hat \eta_\La p_\LA 
+ \frac{m^2_\La}{\hat \eta_\La s}\, p_\LB
\;\;.
\\
\end{split}
\end{equation}
The momentum of the final state spectator partons changes in order to make some momentum available to allow both $p_\La$ and $\hat p_\La$ to be on shell with zero transverse momenta. We denote the total momentum of the final state partons before the splitting by $Q = p_\La + p_\Lb$ and after the splitting by $\hat Q = \hat p_\La + p_\Lb$. In the splitting function, we make use of a lightlike vector $n_\La$ in the plane of $p_\La$ and $Q$. With a convenient choice of normalization, $n_\La = p_\LB$. 

In the following formulas, it will be convenient to define $P_\La = \hat p_\La - \hat p_\mpone$.
 
Using the definition Eq.~(\ref{eq:barwlldef}) with the splitting amplitudes $v_\La$ from Table 1 of Ref.~\cite{NSshower}, we write the spin averaged splitting function as
\begin{equation}
\begin{split}
\overline W_{\La\La} ={}& \frac{4\pi \alpha_\Ls}{2(p_\La\cdot p_\LB)^2}\,
\frac{1}{(2\,\hat p_\La\cdot \hat p_\mpone)^2}\
D_{\mu\nu}(\hat p_\mpone,\hat Q)
\\&\times
\frac{1}{4}{\rm Tr}\left[
(\s{\hat p}_\La + m_\La)\gamma^\mu (\s{P}_\La + m_\La)\s{n}_\La
(\s{p}_\La + m_\La)\s{n}_\La(\s{P}_\La+ m_\La)\gamma^\nu
\right]
\;\;.
\end{split}
\end{equation}
Here $D_{\mu\nu}$ is given in Eq.~(\ref{eq:Dmunu}).

The spin averaged splitting function can be simplified. Let us adopt the notation
\begin{equation}
\begin{split}
\Phi_a = \frac{m_\La^2}{\hat \eta_\La \eta_\Lb s}
\;\;,
\hskip 2 cm
\Phi_b =  \frac{m_\Lb^2}{\hat \eta_\La \eta_\Lb s}
\;\;.
\end{split}
\end{equation}
Then the result can conveniently be displayed in terms of the dimensionless function 
\begin{equation}
\label{eq:FdefISqqg}
F \equiv \frac{\hat p_\La\cdot \hat p_\mpone}{4\pi\as}\
\overline W_{\La\La}
\;\;.
\end{equation}
The result is
\begin{equation}
\begin{split}
\label{eq:qqgISsplit}
F ={}&
F_{\rm eikonal}
+
\frac{\hat p_\mpone\cdot p_\LB}{p_\La\cdot p_\LB}
+
\frac{\Phi_\La\Phi_\Lb\, 
 (\hat \eta_\La - \eta_\La)^2}
{(1 - \Phi_\La \Phi_\Lb)\,\eta_\La^2 }\,
F_{\rm eikonal}
 \;\;.
\end{split}
\end{equation}
Here the first term is the simple eikonal approximation for soft gluon emission,
\begin{equation}
\label{eq:FeikonalIS}
F_{\rm eikonal} = 
\frac{\hat p_\La \cdot D(\hat p_\mpone,\hat Q) \cdot \hat p_\La}
{\hat p_\La\cdot \hat p_\mpone}
\;\;.
\end{equation}
The second term is present in the case of massless or massive quarks and is manifestly positive. The third term is present only if $m_\La$ and $m_\Lb$ are both non-zero. It is likewise manifestly positive as long as $\Phi_\La \Phi_\Lb < 1$. Thus, as for the final state splitting analyzed in the previous section,
\begin{equation}
F - F_{\rm eikonal} \ge 0
\;\;.
\end{equation}

\FIGURE{
\centerline{\includegraphics[width = 10 cm]{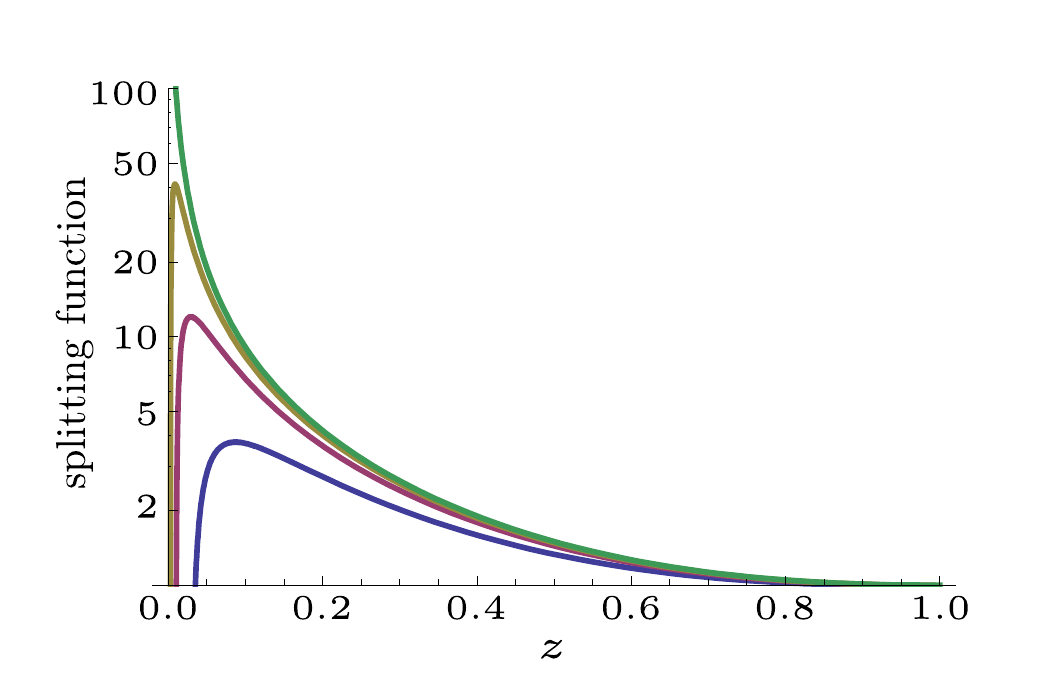}}
\caption{The spin averaged splitting function $(\eta_\La/\hat \eta_\La) F$, with $F$ defined in Eq.~(\ref{eq:FdefISqqg}), for an initial state $q \to q + {\rm g}$ splitting, plotted versus the momentum fraction $z$ of the gluon, as defined in Eq.~(\ref{eq:zISdef}). All partons are taken to be massless. The four curves are for, from bottom to top, $y = 0.03, 0.01, 0.001, 0$. For $y = 0$, the result is the Altarelli-Parisi function, $[1 + (1-z)^2]/z$.}
\label{fig:qqgISplot}
} 

Let us now specialize to $m_\La = m_\Lb = 0$ and examine the behavior of $F$ in more detail. We define a virtuality variable
\begin{equation}
y = \frac{\hat p_\mpone\cdot \hat p_\La}{\hat p_\La \cdot p_\Lb}
\end{equation}
and a variable representing the momentum fraction of the daughter gluon
\begin{equation}
z = \frac{\hat p_\mpone \cdot (p_\Lb - \hat p_\mpone)}
{\hat p_\La \cdot (p_\Lb - \hat p_\mpone)}
\;\;.
\end{equation}
We can write $z$ in a different form by using the kinematic relation that is used to define the momentum mapping ${\cal R}_\La$, $(p_\La + p_\Lb)^2 = (\hat p_\La + p_\Lb - \hat p_\mpone)^2$, which is equivalent to
\begin{equation}
\label{eq:ISkinematic}
\hat p_\La \cdot \hat p_\mpone =
(P_\La - p_\La)\cdot p_\Lb
\;\;.
\end{equation}
This relation gives
\begin{equation}
\label{eq:zISdef}
z = \frac{x-y}{1-y}
\;\;,
\end{equation}
where
\begin{equation}
x = 1 - \frac{\eta_\La}{\hat \eta_\La}
\;\;.
\end{equation}
Note that $z$ and $x$ are equivalent when $y = 0$ but $z$ varies in the range $0 < z < 1$. The inverse relation is
\begin{equation}
x = z + y (1-z)
\;\;.
\end{equation}
A simple calculation gives
\begin{equation}
F = F_{\rm eikonal} + \frac{x}{1-x} - y\;\;,
\end{equation}
where
\begin{equation}
F_{\rm eikonal} = \frac{2}{x}- \frac{2y}{x^2}
\;\;.
\end{equation}
As expected, $(1-x) F = (\eta_\La/\hat \eta_\La) F$ approaches the Altarelli-Parisi splitting function, $[1 + (1-z)^2]/z$ as $y \to 0$. The approach to the limit is depicted in Fig.~\ref{fig:qqgISplot}.

\subsection{Final state ${\rm g} \to {\rm g} + {\rm g}$ splitting}
\label{sec:gggFS}

\FIGURE{
\centerline{\includegraphics[width = 6 cm]{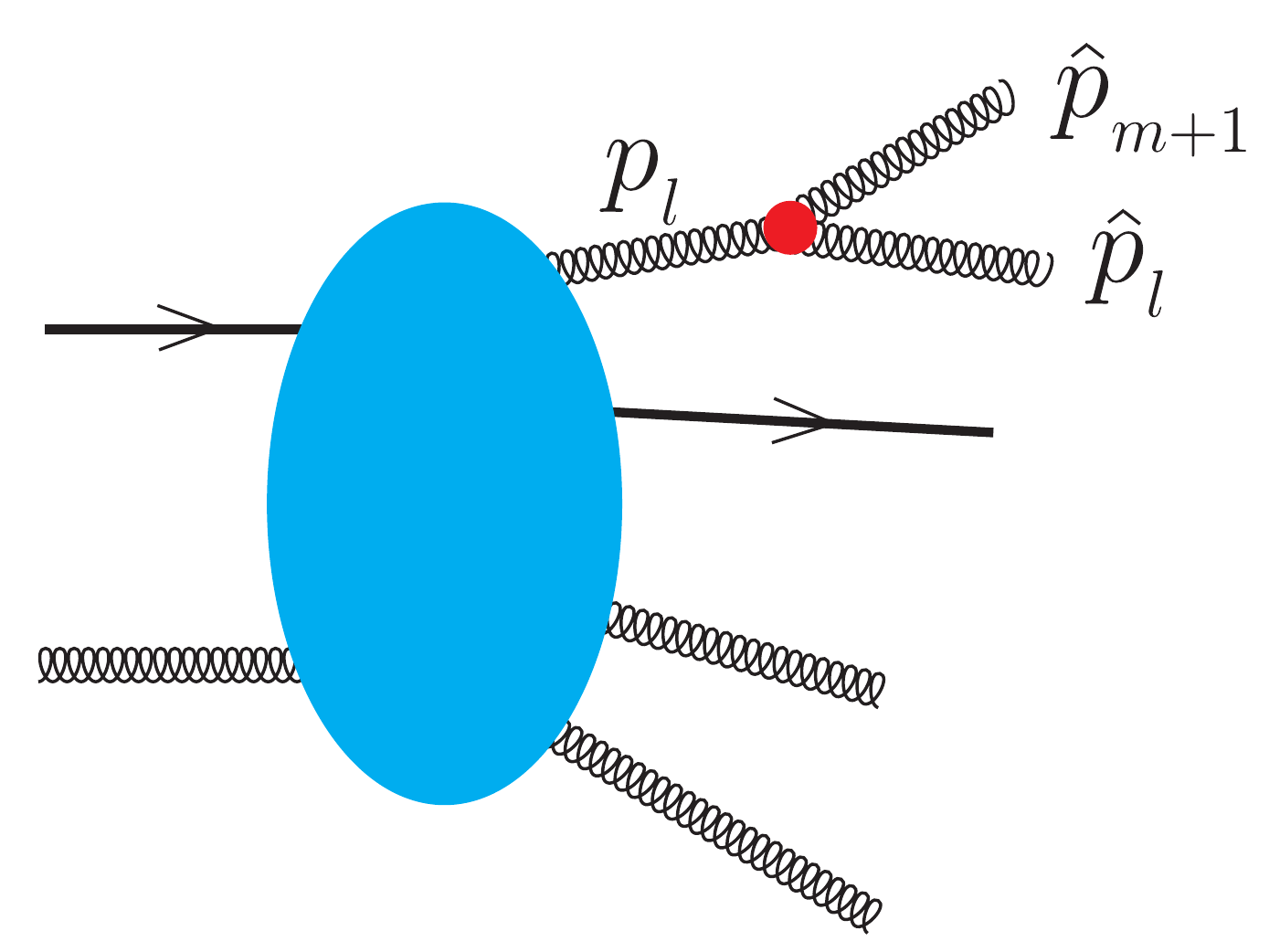}}
\caption{A final state ${\rm g} \to {\rm g} + {\rm g}$ splitting.}
\label{fig:gggFSsplit}
} 

Next we consider a final state $\mathrm g \to \mathrm g \mathrm g$ splitting, as illustrated in Fig.~\ref{fig:gggFSsplit}. According Ref.~\cite{NSshower}, the splitting amplitude is built from the ggg vertex,
\begin{equation}
\label{eq:vgg}
v^{\alpha \beta \gamma}(p_a, p_b, p_c)
= v_1^{\alpha \beta \gamma}(p_a, p_b, p_c)
+ v_2^{\alpha \beta \gamma}(p_a, p_b, p_c)
+ v_3^{\alpha \beta \gamma}(p_a, p_b, p_c)
\;\;,
\end{equation}
where
\begin{equation}
\begin{split}
\label{eq:vgg123}
v_1^{\alpha \beta \gamma}(p_a, p_b, p_c)
={}& g^{\alpha\beta} (p_a - p_b)^\gamma
\;\;,
\\
v_2^{\alpha \beta \gamma}(p_a, p_b, p_c)
={}& g^{\beta\gamma} (p_b - p_c)^\alpha
\;\;,
\\
v_3^{\alpha \beta \gamma}(p_a, p_b, p_c)
={}& g^{\gamma\alpha} (p_c - p_a)^\beta
\;\;.
\end{split}
\end{equation}
We use $v^{\alpha \beta \gamma}$ to define the splitting amplitude
\begin{equation}
\begin{split}
\label{eq:VggF}
v_l(\{\hat p, \hat f\}_{m+1},&\hat s_{m+1},\hat s_{l},s_l)
\\ & =
\frac{\sqrt{4\pi\as}}{2 \hat p_{m+1}\!\cdot\! \hat p_l}\, 
\varepsilon_{\alpha}(\hat p_{m+1}, \hat s_{m+1};\hat Q)^*
\varepsilon_{\beta}(\hat p_{l}, \hat s_l;\hat Q)^*
\varepsilon^{\nu}(p_{l}, s_l;\hat Q)
\\&\quad\times
v^{\alpha \beta \gamma}(\hat p_{m+1},\hat p_l,-\hat p_{m+1}-\hat p_l)\,
D_{\gamma\nu}(\hat p_l + \hat p_{m+1};n_l)
\;\;.
\end{split}
\end{equation}
We have the ggg vertex, polarization vectors for the external particles, and a propagator $D/(2 \hat p_{m+1}\!\cdot\! \hat p_l)$ for the off-shell gluon. The numerator $D_{\gamma\nu}(\hat p_l + \hat p_{m+1};n_l)$ projects on to the physical polarization states for the off-shell gluon,
\begin{equation}
\label{eq:Dmunulightlike}
D_{\mu\nu}(P,n_l) =
- g_{\mu\nu} 
+ \frac{P^\mu n_l^\nu + n_l^\mu P^\nu}
{\hat P\cdot n_l}
\;\;.
\end{equation}
Here $n_l$ is a lightlike vector in the plane of $p_l$ and $\hat Q$, defined as in Eq.~(\ref{eq:nldef}). Then $n_l^\gamma D_{\gamma\nu} = 0$.

Following Ref.~\cite{NSshower}, we define the spin averaged splitting function using Eq.~(\ref{eq:barwlldef}),
\begin{equation}
\begin{split}
\label{eq:wlldef}
\overline W_{ll}={}&
\frac{1}{2}
\Bigg( \frac{1}{2}\sum_{\hat s_l,\hat s_{m+1}, s_l}\Bigg)
\bigg\{
\big|v_l(\{\hat p, \hat f\}_{m+1},\hat s_{m+1},\hat s_{l},s_l)\big|^2
\\ & + 
|v_{2,l}(\{\hat p, \hat f\}_{m+1},\hat s_{m+1},\hat s_{l},s_l)|^2
-
|v_{3,l}(\{\hat p, \hat f\}_{m+1},\hat s_{m+1},\hat s_{l},s_l)|^2\,
\bigg\}
\;\;.
\end{split}
\end{equation}
Here $v_{2,l}$ and $v_{3,l}$ are defined as in Eq.~(\ref{eq:VggF}), but with $v_2^{\alpha \beta \gamma}$ or $v_3^{\alpha \beta \gamma}$, respectively, replacing the full ggg vertex $v^{\alpha \beta \gamma}$. Note first of all the prefactor 1/2, which is a statistical factor for having two identical final state particles in a ${\rm g} \to {\rm g} + {\rm g}$ splitting. This is the factor $S_l$ in Eq.~(\ref{eq:barwlldef}). Then we add $|v_{2,l}|^2 - |v_{3,l}|^2$. This does not change the result when we add this function to the same function with the roles of the two daughter gluons interchanged. With this modification, there is a singularity when daughter gluon $m+1$ becomes soft but not when daughter gluon $l$ becomes soft.

One can evaluate $\overline W_{ll}$ as given in Eq.~(\ref{eq:wlldef}) by using
\begin{equation}
\sum_{s}
\varepsilon_{\mu}(k, s;\hat Q)^*
\varepsilon_{\nu}(k, s;\hat Q)
= D_{\mu\nu}(k,\hat Q)
\;\;,
\end{equation}
where $D_{\mu\nu}(k,\hat Q)$ is defined in Eq.~(\ref{eq:Dmunu}). One might expect a complicated result, but $\overline W_{ll}$ is actually quite simple.
As in previous subsections, we display the result in terms of the dimensionless function 
\begin{equation}
\label{eq:FdefFSggg}
F \equiv \frac{\hat p_l\cdot \hat p_\mpone}{4\pi\as}\
\overline W_{ll}
\;\;.
\end{equation}
The result is
\begin{equation}
\label{FgggFS}
F =
F_{\rm eikonal}
+\frac{(-k_{\perp}^{2})[1 + (1-\Delta)^2]}
{4\,\hat p_{l}\!\cdot\!\hat p_{m+1}}\, 
\;\;.
\end{equation}
Here $F_{\rm eikonal}$ is the standard eikonal function given in Eq.~(\ref{eq:Feikonal}) and
\begin{equation}
\begin{split}
\label{eq:kperpdef}
k_{\perp}^\mu ={}& D(p_{l},\hat Q)^\mu_{\ \nu}\ \hat p_{m+1}^\nu
\;\;,
\\
\Delta ={}& \frac{\hat Q^{2}\,\hat p_{l}\!\cdot\!\hat p_{m+1}}
{\hat p_{l}\!\cdot\!\hat Q\,\hat p_{m+1}\!\cdot\!\hat Q}
\;\;.
\end{split}
\end{equation}
Since $k_\perp$, the part of $\hat p_{m+1}$ orthogonal to $p_l$ and $\hat Q$, is spacelike, we again find
\begin{equation}
F - F_{\rm eikonal} \ge 0
\;\;.
\end{equation}

\FIGURE{
\centerline{\includegraphics[width = 10 cm]{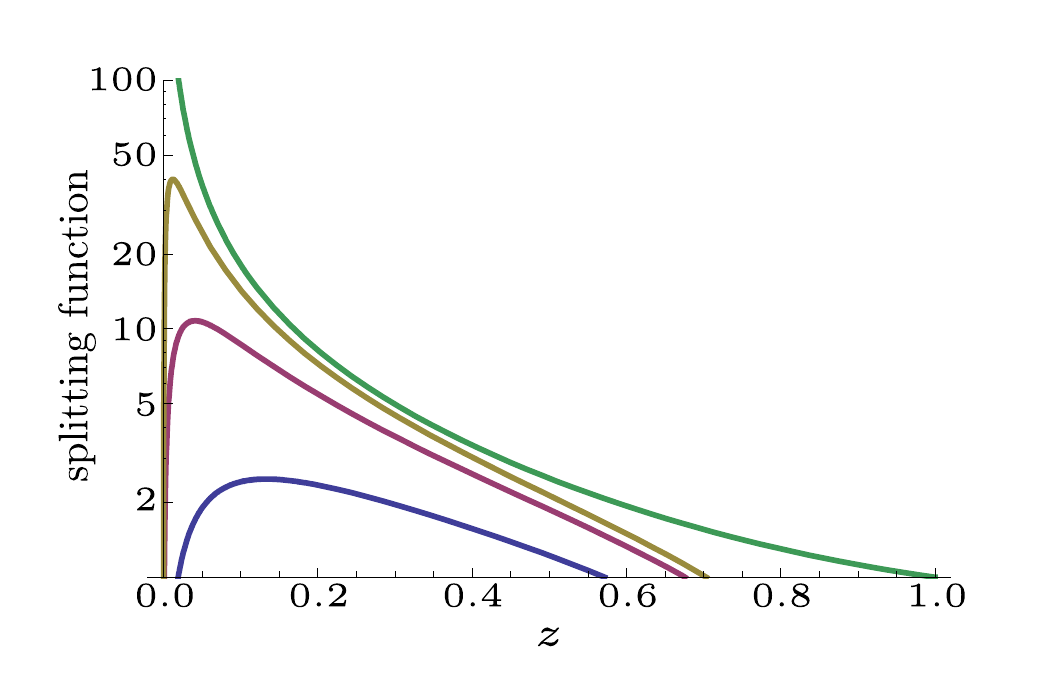}}
\caption{The spin averaged splitting function $F$ defined in Eq.~(\ref{eq:FdefFSggg}) for a final state ${\rm g} \to {\rm g} + {\rm g}$ splitting, plotted versus the momentum fraction $z$ of the gluon, as defined in Eq.~(\ref{eq:zFSdef}). We set $a_l = 4$. The four curves are for, from bottom to top, $y = 0.03, 0.01, 0.001, 0$. For $y = 0$, the result is $2(1-z)/z + z(1-z)$. The sum of this and the same function with $z \to 1-z$ is the standard Altarelli-Parisi function, $2(1-z)/z + 2z/(1-z) + 2z(1-z)$.}
\label{fig:gggFSplot}
} 

We can evaluate $F$ as a function of the variables $y$ and $z$ and the parameter $a_l$, defined as for a final state quark splitting in Eqs.~(\ref{eq:hdefs}) and (\ref{eq:zFSdef}). We find
\begin{equation}
F = F_{\rm eikonal} + \frac{1 + (1-\Delta)^2}{2}\ z (1-z)
\;\;.
\end{equation}
Here $F_{\rm eikonal}$ was given in terms of $z$ and $y$ in Eq.~(\ref{eq:FeikFS}) and
\begin{equation}
\begin{split}
\Delta ={}& \frac{2 a_l y}{x(1-x)(1+y)^2}
\;\;,
\end{split}
\end{equation}
where the auxiliary momentum fraction $x$ was given in terms of $z$ and $y$ in Eq.~(\ref{eq:xdefFS}).

For $y \to 0$, $F$ becomes
\begin{equation}
F \to f(z) \equiv \frac{2(1-z)}{z} + z(1-z)
\;\;.
\end{equation}
The standard Altarelli-Parisi function,
\begin{equation}
f_{\rm AP}(z) = 2\left[
\frac{1-z}{z} + \frac{z}{1-z} + z(1-z)
\right]
\;\;,
\end{equation}
is $f(z) + f(1-z)$. Recall from Eq.~(\ref{eq:wlldef}) that we broke the symmetry in a ${\rm g} \to {\rm g} + {\rm g}$ splitting in such a way that there is a leading singularity for gluon $m+1$ becomming soft but not for gluon $l$ becoming soft. We could have accomplished the same end by using the full ggg vertex but multiplying the splitting function by $\theta(z<1/2)$. Had we done that, the small $y$ limit of $F$ would have been $f(z) = f_{\rm AP}(z)\, \theta(z<1/2)$. This would also give $f(z) + f(1-z) = f_{\rm AP}(z)$.

The full function $F(z,y,a_l)$ approaches $f(z)$ as $y \to 0$, as illustrated in Fig.~\ref{fig:gggFSplot}.

\subsection{Initial state ${\rm g} \to {\rm g} + {\rm g}$ splitting}
\label{sec:gggIS}

\FIGURE{
\centerline{\includegraphics[width = 7.5 cm]{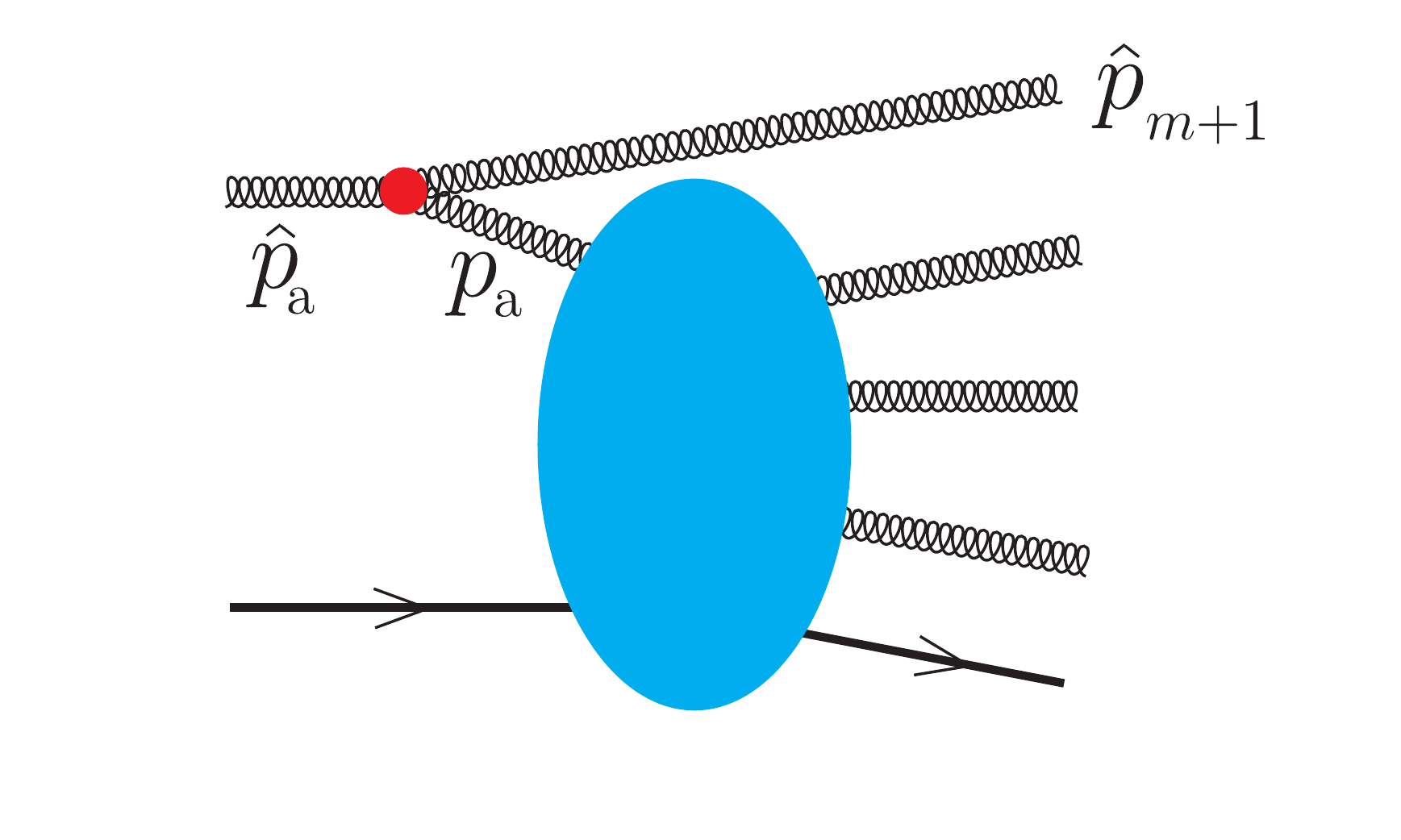}}
\caption{An initial state ${\rm g} \to {\rm g} + {\rm g}$ splitting.}
\label{fig:gggISsplit}
} 

We now consider an initial state $\mathrm g \to \mathrm g \mathrm g$ splitting, as illustrated in Fig.~\ref{fig:gggISsplit}. According Ref.~\cite{NSshower}, the splitting amplitude is again built from the ggg vertex, $v^{\alpha \beta \gamma}$, Eq.~(\ref{eq:vgg}). We use $v^{\alpha \beta \gamma}$ to define the splitting amplitude for the splitting of one of the initial state partons, say parton ``a,''
\begin{equation}
\begin{split}
\label{eq:VggI}
v_\La(\{\hat p, \hat f\}_{m+1},&\hat s_{m+1},\hat s_{l},s_l)
\\&=
-
\frac{\sqrt{4\pi\as}}{2 \hat p_{m+1}\!\cdot\! \hat p_\La}\, 
\varepsilon_{\alpha}(\hat p_{m+1}, \hat s_{m+1};\hat  Q)^*
\varepsilon_{\beta}(\hat p_{\La},\hat s_\La; \hat Q)
\varepsilon^{\nu}(p_{\La}, s_\La; \hat Q)^*
\\&\quad\times
v^{\alpha \beta \gamma}(\hat p_{m+1}, -\hat p_\La, \hat p_\La - \hat p_{m+1})\,
D_{\gamma\nu}(\hat p_\La - \hat p_{m+1};n_\La)
\;\;.
\end{split}
\end{equation}
We have the ggg vertex, polarization vectors for the external particles, and a propagator $D/(2 \hat p_{m+1}\!\cdot\! \hat p_\La)$ for the off-shell gluon. The numerator $D_{\gamma\nu}(\hat p_\La - \hat p_{m+1};n_\La)$ projects on to the physical polarization states for the off-shell gluon. It is defined using Eq.~(\ref{eq:Dmunulightlike}), with the lightlike vector $n_\La = p_\LB$. Following Ref.~\cite{NSshower}, we use Eq.~(\ref{eq:barwlldef}) to define the spin averaged splitting function from the square of $v_\La$,
\begin{equation}
\begin{split}
\label{eq:wlldefgggIS}
\overline W_{\La\La}(\{\hat f,\hat p\}_{m+1})={}&
\frac{1}{2}\sum_{\hat s_\La,\hat s_{m+1}, s_\La}\
\big|v_\La(\{\hat p, \hat f\}_{m+1},\hat s_{m+1},\hat s_{\La},s_\La)\big|^2
\;\;.
\end{split}
\end{equation}

Remarkably, $\overline W_{\La\La}$ is rather simple. As in previous subsections, we display the result in terms of the dimensionless function 
\begin{equation}
\label{eq:FdefISggg}
F \equiv \frac{\hat p_\La\cdot \hat p_\mpone}{4\pi\as}\
\overline W_{\La\La}
\;\;.
\end{equation}
The result is
\begin{equation}
\begin{split}
\label{Fgini}
F ={}& F_{\rm eikonal}
+\frac{- k_{\perp}^{2}}{\hat p_{\La}\!\cdot\!\hat p_{m+1}}\,
\left[\left(\frac{\hat p_{m+1}\!\cdot\!n_{\La}}
{(\hat p_\La - \hat p_{m+1})\!\cdot\!n_{\La}}\right)^{2}
+\frac{\hat p_\La\!\cdot\!n_\La\ \hat p_{m+1}\!\cdot\!\hat Q
+ \hat p_{m+1}\!\cdot\!n_{\La}\ \hat p_{\La}\!\cdot\!\hat Q
}
{(\hat p_\La - \hat p_{m+1})\!\cdot\!n_{\La}\ \hat p_{m+1}\!\cdot\!\hat Q}
\right]
\;\;.
\end{split}
\end{equation}
Here $F_{\rm eikonal}$ is the eikonal function, Eq.~(\ref{eq:FeikonalIS}), and $k_{\perp}^\mu = D(p_{\La},\hat Q)^\mu_{\ \nu}\ \hat p_{m+1}^\nu$ as in Eq.~(\ref{eq:kperpdef}). Examination of Eq.~(\ref{Fgini}) shows that, as in the previous cases,
\begin{equation}
F  - F_{\rm eikonal} \ge 0
\;\;.
\end{equation}
To see this, one needs to know that splitting kinematics ensures that $(\hat p_\La - \hat p_{m+1})\!\cdot\!n_{\La} > 0$. We note that the splitting kinematics allows non-zero parton masses, although the gluon that splits here is massless.

Let us look at this assuming massless partons and using the splitting variables $y$, $z$ and $x = z + y (1-z)$ defined in Sec.~(\ref{sec:qqgIS}). A straightforward calculation gives
\begin{equation}
F = \frac{2}{x} - \frac{2y}{x^2}
+{2 (1-y) z}
\left[
\left(\frac{(1-y)z}{1 - (1-y)z}\right)^2
+ \frac{2 z (1-y) + y}{z(1-z)(1-y)^2 + y}
\right]
\;\;.
\end{equation}
As expected, $(1-x) F = (\eta_\La/\hat \eta_\La) F$ approaches the Altarelli-Parisi splitting function, $2z/(1-z) + 2(1-z)/z + 2 z(1-z)$ as $y \to 0$. The approach to the limit is depicted in Fig.~\ref{fig:gggISplot}.

\FIGURE{
\centerline{\includegraphics[width = 10 cm]{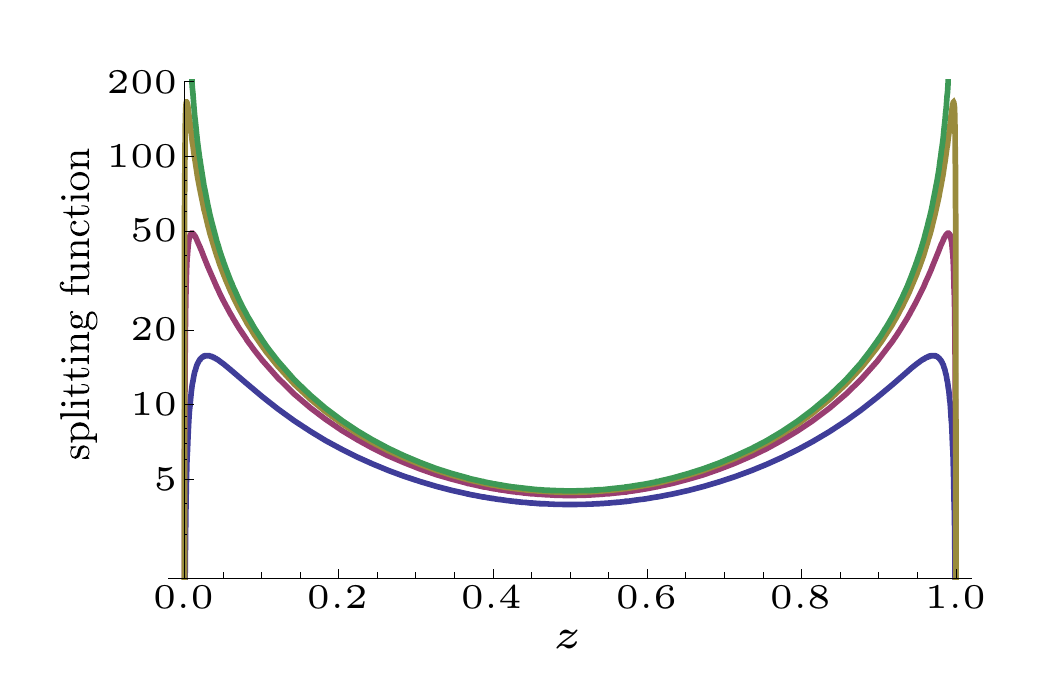}}
\caption{
The spin averaged splitting function $(\eta_\La/\hat \eta_\La) F$, with $F$ defined in Eq.~(\ref{eq:FdefISggg}), for an initial state ${\rm g} \to {\rm g} + {\rm g}$ splitting, plotted versus the momentum fraction $z$ of the gluon, as defined in Eq.~(\ref{eq:zFSdef}). The four curves are for, from bottom to top, $y = 0.03, 0.01, 0.001, 0$. For $y = 0$, the result is the standard Altarelli-Parisi function, $2(1-z)/z + 2z/(1-z) + 2z(1-z)$.}
\label{fig:gggISplot}
} 

\subsection{Other cases}
\label{sec:OtherDirect}

We have covered the cases of quark or gluon splittings in which a daughter gluon enters the final state. There is also the possibility of an antiquark splitting replacing a quark spitting, but, because of charge conjugation invariance, these are essentially the same as the quark splitting cases. There are also cases in which no daughter gluon enters the final state: final state and initial state $\mathrm g \to q + \bar q$ and initial state $q \to q + \mathrm g$ and $\bar q \to \bar q + \mathrm g$ in which the gluon enters the hard scattering and the quark or antiquark enters the final state. The spin averaged splitting functions for these cases are manifestly positive. In these cases, there is no leading singularity when a final state daughter parton becomes soft, so we do not need to consider soft gluon singularities. We list the results for these cases in Appendix~\ref{app:OtherSplittings}.

\section{Interference diagrams}
\label{sec:interference}

\FIGURE{
\centerline{\includegraphics[width = 13 cm]{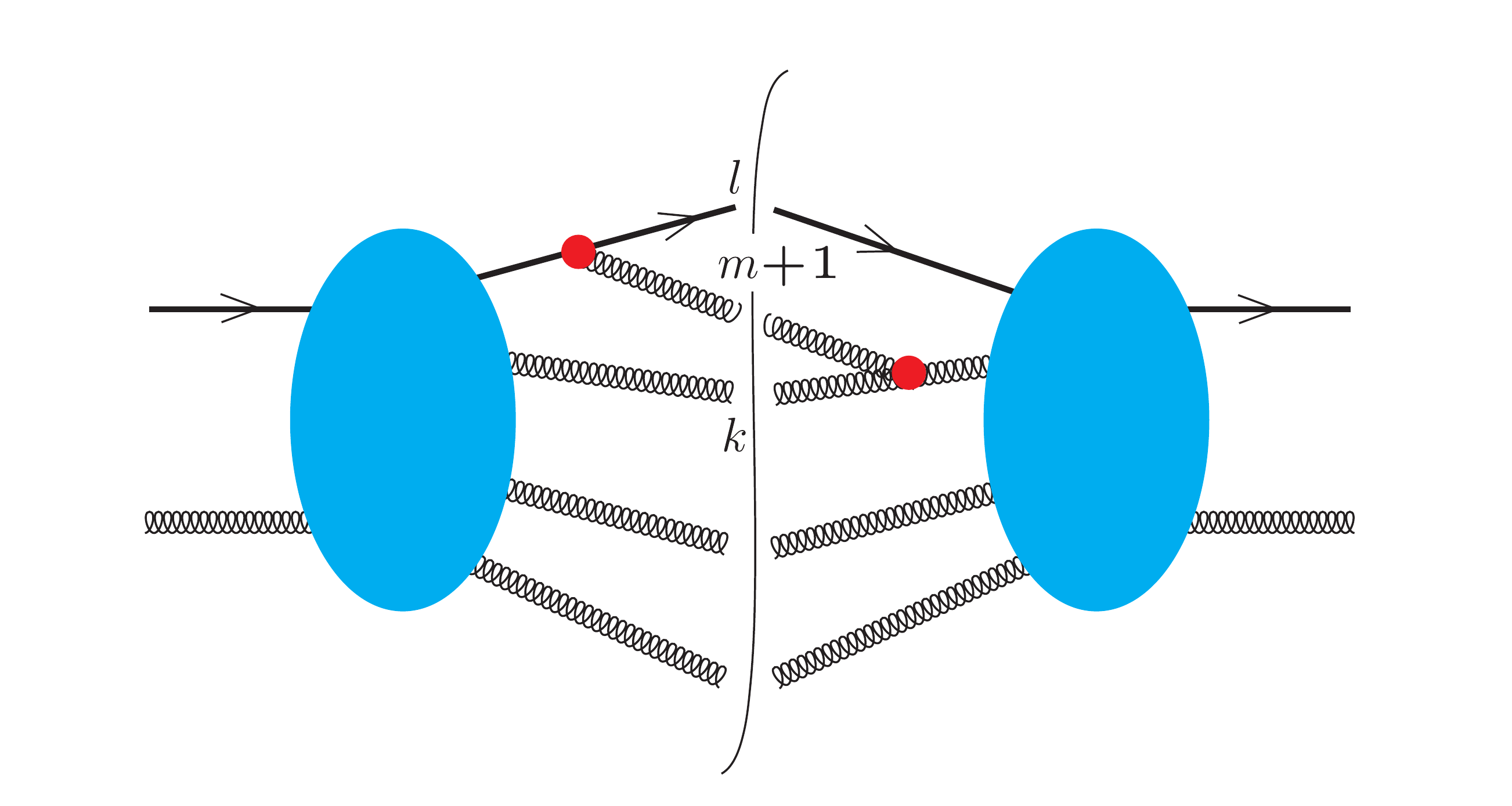}}
\caption{
An interference diagram. A gluon, labelled $m+1$, is emitted from parton $l$ in the amplitude and from parton $k$ in the complex conjugate amplitude. This diagram has a leading singularity when the gluon is soft.}
\label{fig:interference}
} 

We have analyzed the spin averaged splitting functions $\overline W_{ll}$, which correspond to the squared amplitude for a parton $l$ to split into daughter partons with labels $l$ and $m+1$. Now we need to consider interference diagrams, such as the diagram illustrated in Fig.~\ref{fig:interference}. In the amplitude, parton $l$ can change into a daughter parton with label $l$ by emitting a gluon with label $m+1$. In the complex conjugate amplitude, parton $k$ can change into a daughter parton with label $k$ by emitting a gluon with label $m+1$. If we were to temporarily ignore questions about how to define the kinematics and were to use the splitting amplitudes $v_l$ and $v_k$ for this, the corresponding contribution to the splitting function would be
\begin{equation}
v_l(\{\hat p, \hat f\}_{m+1},\hat s_{m+1},\hat s_{l},s_l)\,
\delta_{\hat s_k,s_k}\
v_k(\{\hat p, \hat f\}_{m+1},\hat s_{m+1}',\hat s_{k}',s_k')^*\,
\delta_{\hat s_l',s_l'}
\;\;.
\end{equation}
This function is singular when gluon $m+1$ is soft, $\hat p_{m+1} \to 0$. However it does not have a leading singularity when gluon $m+1$ is collinear with parton $l$ or parton $k$. For this reason, we can use a simple eikonal approximation to the splitting amplitude,
\begin{equation}
v_l^{\rm soft}(\{\hat p, \hat f\}_{m+1},\hat s_{m+1},\hat s_{l},s_l)
= \sqrt{4\pi\as}\, \delta_{\hat s_l,s_l}\,
\frac{
\varepsilon(\hat p_{m+1}, \hat s_{m+1};Q)^*
\!\cdot\! \hat p_l}
{\hat p_{m+1}\!\cdot\! \hat p_l}
\;\;,
\end{equation}
if parton $m+1$ is a gluon, with $v_l^{\rm soft} = 0$ otherwise.

Making the eikonal approximation, the splitting function is
\begin{equation}
\label{eq:softapprox0}
W_{lk} = 
v_l^{\rm soft} (v_k^{\rm soft})^*\
\delta_{\hat s_l',s_l'}\ \delta_{\hat s_k,s_k}
\;\;.
\end{equation}
This function gives the dependence of the splitting operator on momentum and spin. In Ref.~\cite{NSshower} there is a separate factor that gives the color dependence. This factor is an operator on the color space that we can call $t^{\dagger}_l \otimes t^{}_k$, where $t^{\dagger}_l$ is the operator in Eq.~(\ref{eq:splittingamplitudestructure}) that inserts the proper color matrix into line $l$ in the amplitude and $t^{}_k$ inserts the proper color matrix into line $k$ in the complex conjugate amplitude.\footnote{In Ref.~\cite{NSshower}, we write $t^\dagger_l(f_l \to f_l + {\rm g})$ for the operator that we here call just $t^\dagger_l$ and we denote the operator $t^{\dagger}_l \otimes t^{}_k$ by ${\cal G}(l,k)$.} We do not yet make any approximations with respect to color. In Sec.~\ref{sec:color}, we will make the approximation of keeping only the leading color conributions.

There is an ambiguity with the prescription (\ref{eq:softapprox0}). The functions in $W_{lk}$ are defined in terms of daughter parton momenta and flavors, $\{\hat p, \hat f\}_{m+1}$. However, we want to define $\{\hat p, \hat f\}_{m+1}$ from the momenta and flavors $\{p,f\}_{m}$ before the splitting together with the splitting variables $\{\zeta_{\rm p}, \zeta_{\rm f}\}$. We need to specify what relation to use. One way is to use the kinematic functions that we use for the splitting of parton $l$ into a daughter parton $l$ and the gluon $m+1$,
$
\{\hat p, \hat f\}_{m+1} = R_l(\{p,f\}_m, \{\zeta_{\rm p},\zeta_{\rm f}\}).
$
With this mapping, we define a function $W^{(l)}_{lk}$ of the $\{p,f\}_{m}$ and the splitting variables. Alternatively, we could use the kinematic functions, $R_k$, that we use for the splitting of parton $k$ into a daughter parton $k$ and the gluon $m+1$. With this momentum mapping, we define a function $W^{(k)}_{lk}$ of the $\{p,f\}_{m}$ and the splitting variables. Instead of using one or the other of these possibilities, we average over them. We use $W^{(l)}_{lk}$ with weight $A_{lk}$ and $W^{(k)}_{lk}$ with weight $A_{kl}$. In Ref.~\cite{NSshower}, we let the weight functions take the default value $A_{lk} = A_{kl} = 1/2$. This choice is certainly conceptually simple. However, we can obtain spin-summed splitting functions that have nicer properties if we define the weights as certain functions $A_{lk}(\{\hat p\}_{m+1})$ and $A_{kl}(\{\hat p\}_{m+1})$ of the momenta. It is simplest to specify the functional forms of the weight functions using the momenta $\{\hat p\}_{m+1}$ after splitting. The momenta after splitting are to be determined by the mapping ${\cal R}_l$ for $A_{lk}$ and by the mapping ${\cal R}_k$ for $A_{kl}$.\footnote{This is expressed most precisely using the operator language of Eq.~(8.26) of Ref.~\cite{NSshower}.} The weight functions are non-negative and obey $A_{lk}(\{\hat p\}_{m+1}) + A_{kl}(\{\hat p\}_{m+1}) = 1$ at fixed momenta $\{\hat p\}_{m+1}$. The relation $A_{lk} + A_{kl} \approx 1$ then holds at fixed $\{p,f\}_{m}$ and splitting variables. This approximate relation becomes exact in the limit of an infinitely soft splitting, for which the mappings ${\cal R}_l$ and ${\cal R}_k$ become identical.

With the choice of momentum mappings determined by $A_{lk}$ and $A_{kl}$, the net splitting function, including the color factor, summed over the two graphs arising from interference of soft gluons emitted from partons $l$ and $k$, is
\begin{equation}
\begin{split}
& \left[
A_{lk} W^{(l)}_{lk}
+
A_{kl} W^{(k)}_{lk}
\right]  \, t^{\dagger}_l \otimes t^{}_k
+
\left[
A_{kl} W^{(k)}_{kl}
+
A_{lk} W^{(l)}_{kl}
\right] \, t^{\dagger}_k \otimes t^{}_l
=
\\&\qquad
A_{lk} 
\left[
W^{(l)}_{lk}\, t^{\dagger}_l \otimes t^{}_k
+
W^{(l)}_{kl}\, t^{\dagger}_k \otimes t^{}_l
\right]\
+
A_{kl} 
\left[
W^{(k)}_{lk}\, t^{\dagger}_l \otimes t^{}_k
+
W^{(k)}_{kl}\, t^{\dagger}_k \otimes t^{}_l
\right]
\;\;.
\end{split}
\end{equation}

We will see in the following section that we obtain spin-summed splitting functions that have nice properties if we define $A_{lk}$ as a ratio,
\begin{equation}
\label{eq:Alkdef}
A_{lk}(\{\hat p\}_{m+1})
= \frac{B_{lk}(\{\hat p\}_{m+1})}
{B_{lk}(\{\hat p\}_{m+1})
+B_{kl}(\{\hat p\}_{m+1})}
\;\;,
\end{equation}
where
\begin{equation}
\label{eq:Blkdef}
B_{lk}(\{\hat p\}_{m+1}) = 
\frac{\hat p_{m+1} \!\cdot\! \hat p_k}{\hat p_{m+1} \!\cdot\! \hat p_l}\
\hat p_l \!\cdot\! D(\hat p_{m+1},\hat Q) \!\cdot\! \hat p_l 
\;\;.
\end{equation}
Here $D(\hat p_{m+1}, \hat Q)$ is the transverse projection tensor defined in Eq.~(\ref{eq:Dmunu}). 

\section{Spin-averaged interference graph splitting functions}
\label{sec:interferenceSFs}

The part of the soft splitting function representing $l$-$k$ interference that is associated with the kinematic mapping ${\cal R}_l$ is
\begin{equation}
A_{lk} 
\left[
W^{(l)}_{lk}\, t^{\dagger}_l \otimes t^{}_k
+
W^{(l)}_{kl}\, t^{\dagger}_k \otimes t^{}_l
\right]
\;\;.
\end{equation}
We now make the approximation of setting $s' = s$ for each parton, summing over the daughter parton spins, and averaging over the mother parton spins. The sum over spins of $W^{(l)}_{lk}$ is the same as the sum over spins of $W^{(l)}_{kl}$. Thus the spin averaged splitting function, including the color factor, becomes\footnote{The function $\overline W_{lk}$ here equals the product $2 A_{lk}\overline w_{lk}$ of functions in Ref.~\cite{NSshower}.}
\begin{equation}
\frac{1}{2}
\left[t^{\dagger}_l \otimes t^{}_k + t^{\dagger}_k \otimes t^{}_l\right]\,
\overline W_{lk}
\;\;,
\end{equation}
where
\begin{equation}
\overline W_{lk} = \frac{1}{4}\sum_{ s_l,\hat s_l,s_k,\hat s_k,\hat s_{m+1}}
A_{lk} 
\left[
W^{(l)}_{lk}
+
W^{(l)}_{kl}
\right]_{\{s'\} = \{s\}}
\;\;.
\end{equation}
Here we have used the notation $\{s'\} = \{s\}$ to indicate the instruction to set $s'_l = s_l$, $\hat s'_l = \hat s_l$, $s'_k = s_k$, $\hat s'_k = \hat s_k$, and $\hat s'_{m+1} = \hat s_{m+1}$. The structure of $\overline W_{lk}$ is quite simple,
\begin{equation}
\begin{split}
\overline W_{lk} = {}&
4\pi\as\ 2 A_{lk}\
\frac{
 \hat p_l\cdot D(\hat p_{m+1};\hat Q) \cdot  \hat p_k}
{\hat p_{m+1}\!\cdot\! \hat p_l\ \hat p_{m+1}\!\cdot\! \hat p_k}
\;\;.
\end{split}
\end{equation}

We can associate $\overline W_{lk}$ with the splitting of parton $l$, since it uses the kinematic mapping ${\cal R}_l$. Then we are led to consider the relation of $\overline W_{lk}$ to the direct splitting function $\overline W_{ll}$. Now, the color factor that multiplies $\overline W_{ll}$ is $t^{\dagger}_l \otimes t^{}_l$. However, as discussed in Ref.~\cite{NSshower}, the invariance of the matrix element under color rotations implies that 
\begin{equation}
t^{\dagger}_l \otimes t^{}_l
= -\sum_{k \ne l}
\frac{1}{2}
\left[t^{\dagger}_l \otimes t^{}_k + t^{\dagger}_k \otimes t^{}_l\right]
\;\;.
\end{equation}
Thus we can combine the direct and interference graphs to give
\begin{equation}
\label{eq:colorandspin}
\left(
-\frac{1}{2}
\left[t^{\dagger}_l \otimes t^{}_k + t^{\dagger}_k \otimes t^{}_l\right]
\right)\
\left[\overline W_{ll} - \overline W_{lk}
\right]
\;\;.
\end{equation}
We will see in Sec.~\ref{sec:color} that the color factor here is very simple in the leading color limit, essentially amounting to multiplying by $C_{\rm F}$ or zero. We are thus motivated to investigate the coefficient of this color operator, $\overline W_{ll} -\overline W_{lk}$. 

It is useful to break $\overline W_{ll} - \overline W_{lk}$ into two pieces, 
\begin{equation}
\overline W_{ll}
-
\overline W_{lk}
=
(\overline W_{ll}
-
\overline W_{ll}^{\,\rm eikonal})
+
(
\overline W_{ll}^{\,\rm eikonal}
-
\overline W_{lk})
\;\;.
\end{equation}
Here, we recall from Eq.~(\ref{eq:wlleikonal}), 
\begin{equation}
\begin{split}
\overline W_{ll}^{\rm eikonal} = {}&
4\pi\as\ 
\frac{
 \hat p_l\cdot D(\hat p_{m+1};\hat Q) \cdot  \hat p_l}
{(\hat p_{m+1}\!\cdot\! \hat p_l)^2}\
\;\;.
\end{split}
\end{equation}
We have investigated $(\overline W_{ll} - \overline W_{ll}^{\rm eikonal})$ in Sec.~\ref{sec:directSFs} and found that 
\begin{equation}
\overline W_{ll}
-
\overline W_{ll}^{\rm eikonal}
\ge 0
\;\;.
\end{equation}
Thus we should consider $\overline W_{ll}^{\,\rm eikonal} - \overline W_{lk}$. We have
\begin{equation}
\overline W_{ll}^{\,\rm eikonal}
-
\overline W_{lk}
=
\frac{4\pi\as}{\hat p_{m+1}\!\cdot\! \hat p_l}\
\left\{
\frac{
 \hat p_l\cdot D(\hat p_{m+1};\hat Q) \cdot  \hat p_l}
{\hat p_{m+1}\!\cdot\! \hat p_l}
- 2 A_{lk}\
\frac{
\hat p_l\cdot D(\hat p_{m+1};\hat Q) \cdot  \hat p_k}
{\hat p_{m+1}\!\cdot\! \hat p_k}
\right\}
\;\;.
\end{equation}
We can simplify this if we use the definitions (\ref{eq:Alkdef}) of $A_{lk}$ and (\ref{eq:Blkdef}) of $B_{lk}$,
\begin{equation}
\begin{split}
\overline W_{ll}^{\,\rm eikonal}
&-
\overline W_{lk} =
\\ &
\frac{4\pi\as}{\hat p_{m+1} \cdot \hat p_l}\
\frac{1}{\hat p_{m+1} \cdot \hat p_k}
\bigg\{
\frac{\hat p_{m+1} \cdot \hat p_k}{\hat p_{m+1} \cdot \hat p_l}\
{\hat p_l \cdot D \cdot \hat p_l}
-2 A_{lk}\
{\hat p_l \cdot D \cdot \hat p_k}
\bigg\}
\\ 
={}&
\frac{4\pi\as}{\hat p_{m+1} \cdot p_l}\
\frac{1}{\hat p_{m+1} \cdot \hat p_k}
\bigg\{
B_{lk}
-2 A_{lk}\
{\hat p_l \cdot D \cdot \hat p_k}
\bigg\}
\\ 
={}&
\frac{4\pi\as}{\hat p_{m+1} \cdot \hat p_l}\
\frac{A_{lk}}{\hat p_{m+1} \cdot \hat p_k}\
\bigg\{
B_{lk} + B_{kl}
-2\
{\hat p_l \cdot D \cdot \hat p_k}
\bigg\}
\\ 
={}&
\frac{4\pi\as}{\hat p_{m+1} \cdot \hat p_l}\
\frac{A_{lk}}{\hat p_{m+1} \cdot \hat p_k}\
\bigg\{
\frac{\hat p_{m+1} \cdot p_k}{\hat p_{m+1} \cdot p_l}\
{\hat p_l \cdot D \cdot \hat p_l}
+
\frac{\hat p_{m+1} \cdot p_l}{\hat p_{m+1} \cdot p_k}\
{\hat p_k \cdot D \cdot \hat p_k}
\\ &\qquad 
-2\
{\hat p_l \cdot D \cdot \hat p_k}
\bigg\}
\\ 
={}&
\frac{4\pi\as}{\hat p_{m+1} \cdot \hat p_l}\
\frac{A_{lk}}{\hat p_{m+1} \cdot \hat p_k}\
\frac{1}{\hat p_{m+1} \cdot \hat p_l\ \hat p_{m+1} \cdot \hat p_k}\
\big\{
(\hat p_{m+1} \cdot \hat p_k)^2\
{\hat p_l \cdot D \cdot \hat p_l}
\\ &\quad  
+
(\hat p_{m+1} \cdot \hat p_l)^2\
{\hat p_k \cdot D \cdot \hat p_k}
-2\
(\hat p_{m+1} \cdot \hat p_k)(\hat p_{m+1} \cdot \hat p_l)\
{\hat p_l \cdot D \cdot \hat p_k}
\big\}
\\ 
={}&
\frac{4\pi\as}{\hat p_{m+1} \cdot \hat p_l}\
\frac{A_{lk}}{\hat p_{m+1} \cdot \hat p_k}\
\frac{1}{\hat p_{m+1} \cdot \hat p_l\ \hat p_{m+1} \cdot \hat p_k}\
\\&\quad\times
(\hat p_{m+1} \cdot \hat p_k\ \hat p_l 
- \hat p_{m+1} \cdot \hat p_l\ \hat p_k)
\cdot D \cdot 
(\hat p_{m+1} \cdot \hat p_k\ \hat p_l 
- \hat p_{m+1} \cdot \hat p_l\ \hat p_k)
\;\;.
\end{split}
\end{equation}
We can simplify this further by noting that the vector $\hat p_{m+1} \cdot \hat p_k\ \hat p_l - \hat p_{m+1} \cdot \hat p_l\ \hat p_k$ is orthogonal to $\hat p_{m+1}$, so that only the term $-g^{\mu\nu}$ in $D^{\mu\nu}$ contributes. Thus
\begin{equation}
\overline W_{ll}^{\,\rm eikonal}
-
\overline W_{lk} =
\frac{4\pi\as}{\hat p_{m+1} \cdot \hat p_l}\
\frac{A_{lk}}{\hat p_{m+1} \cdot \hat p_k}\
\frac{-(\hat p_{m+1} \cdot \hat p_k\ \hat p_l 
- \hat p_{m+1} \cdot \hat p_l\ \hat p_k)^2}
{\hat p_{m+1} \cdot \hat p_l\ \hat p_{m+1} \cdot \hat p_k}\
\;\;.
\end{equation}
Since the vector $\hat p_{m+1} \cdot p_k\ p_l - \hat p_{m+1} \cdot p_l\ p_k$ is orthogonal to the lightlike vector $\hat p_{m+1}$, it is either lightlike or spacelike. Furthermore, $A_{lk} \ge 0$. Thus
\begin{equation}
\overline W_{ll}^{\,\rm eikonal}
-
\overline W_{lk} \ge 0
\;\;.
\end{equation}
Thus both parts of our splitting function, $\overline W_{ll} - \overline W_{ll}^{\,\rm eikonal}$ and $\overline W_{ll}^{\,\rm eikonal} - \overline W_{lk}$, are non-negative. This means that we can use these functions as probabilities in constructing a parton shower Monte Carlo program without needing separate weight functions. We discuss this further in Sec.~\ref{sec:EvolutionEquation}. 

The analysis so far has allowed partons $l$ and $k$ to have non-zero masses. Let us now consider the case of massless partons, $\hat p_l^2 = \hat p_k^2 = 0$. The massless result can be understood in more detail if we write it in terms of three-vectors in the frame in which $\vec Q = 0$. We define $\vec u_k$, $\vec u_l$, and $\vec u_{m+1}$ to be unit three-vectors in the directions of the space parts of $\hat p_l$, $\hat p_k$, and $\hat p_{m+1}$ respectively. Then
\begin{equation}
\label{eq:softdistfromg}
\overline W_{ll}^{\rm eikonal}
-
\overline W_{lk} =
\frac{4\pi\as\ 2\hat Q^2}
{(\hat Q\!\cdot\!\hat p_{m+1})^2\ (1 - \vec u_{m+1}\!\cdot\! \vec u_l)}\
g(\vec u_{m+1},\vec u_l,\vec u_k)
\;\;,
\end{equation}
where
\begin{equation}
\label{eq:gdef}
g(\vec u_{m+1},\vec u_l,\vec u_k) =
\frac{
(1 + \vec u_{m+1}\!\cdot\! \vec u_l)
(1 - \vec u_{l}\!\cdot\! \vec u_k)
}
{
(1 - \vec u_{m+1}\!\cdot\! \vec u_l)
(1 + \vec u_{m+1}\!\cdot\! \vec u_k)
+
(1 - \vec u_{m+1}\!\cdot\! \vec u_k)
(1 + \vec u_{m+1}\!\cdot\! \vec u_l)
}
\;.
\end{equation}
We can make some comments about this. First, the splitting probability is singular when the angle between $\vec u_{m+1}$ and $\vec u_l$ approaches zero, $(1 - \vec u_{m+1}\cdot \vec u_l) \to 0$. This is the standard collinear singularity, seen in the soft limit. Second, when $(1 - \vec u_{m+1}\cdot \vec u_l) \ll (1 - \vec u_{k}\cdot \vec u_l) \ll 1$, $\overline W_{ll}^{\,\rm eikonal} - \overline W_{lk}$ behaves like $1/(1 - \vec u_{m+1}\cdot \vec u_l)$. If we integrate over the angle of $\vec u_{m+1}$ with a lower cutoff on the angle between $\vec u_{m+1}$ and $\vec u_{l}$, the integral is logarithmically sensitive to the cutoff. Third, when $(1 - \vec u_{k}\cdot \vec u_l) \ll (1 - \vec u_{m+1}\cdot \vec u_l) \ll 1$, $\overline W_{ll}^{\,\rm eikonal} - \overline W_{lk}$ behaves like $1/(1 - \vec u_{m+1}\cdot \vec u_l)^2$. If we were to put an upper cutoff on the angular integration, there would be no logarithmic sensitivity to this cutoff. Thus, only the angle ordered region $(1 - \vec u_{m+1}\cdot \vec u_l) \lesssim (1 - \vec u_{k}\cdot \vec u_l)$ is important in the integral over angles. There is a smooth decrease in the splitting probability when the angle between $\vec u_{m+1}$ and $\vec u_l$ becomes greater than the angle between $\vec u_{k}$ and $\vec u_l$. There is no sharp cutoff.

\FIGURE{
\centerline{\includegraphics[width = 10 cm]{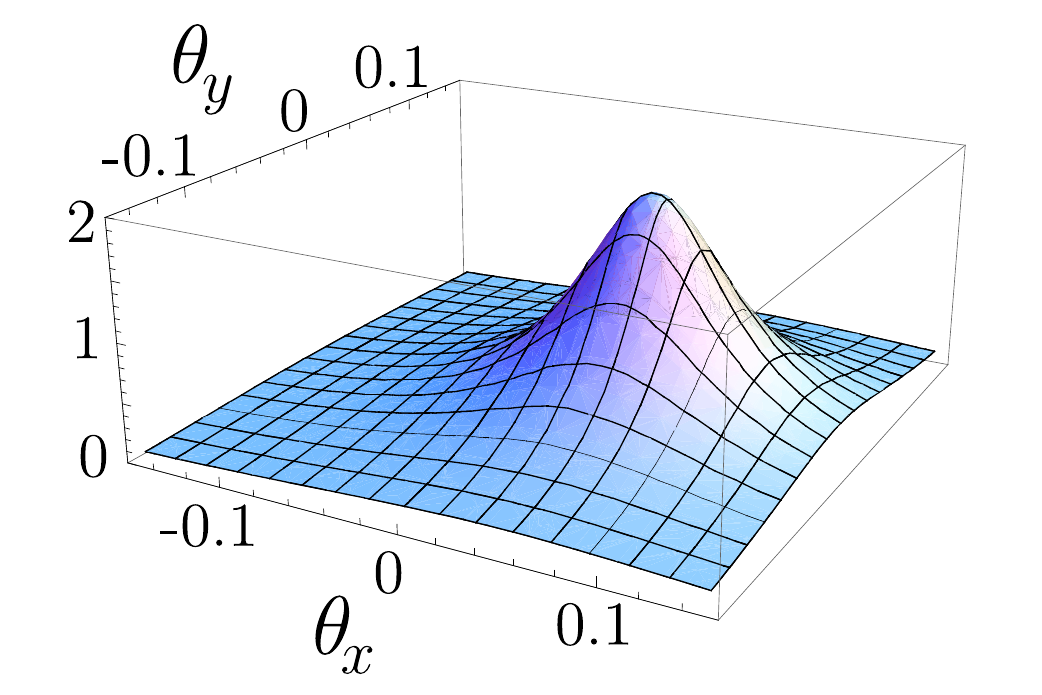}}
\caption{The function $g$ defined in Eq.~(\ref{eq:gdef}) that serves to suppress soft gluon radiation outside of the ``angle ordered'' region. The plot coordinates are $\theta_x = \theta \cos \phi$ and $\theta_y = \theta \sin \phi$, where $\theta,\phi$ are the polar angles of $\vec u_{m+1}$. The vector $\vec u_l$ is at $\theta = 0$ and the vector $\vec u_k$ is at $\theta = 0.1$, $\phi = 0$.}
\label{fig:softplot}
} 

We illustrate this in Fig.~\ref{fig:softplot}. We take the polar angles of $\vec u_{m+1}$ to be $\theta,\phi$ where $\vec u_{l}$ is along the $\theta = 0$ axis. We choose $\vec u_k$ to have polar angles $\theta_k = 0.1$ and $\phi_k = 0$. Then we plot $g(\vec u_{m+1},\vec u_l,\vec u_k)$ versus $\theta_x = \theta \cos \phi$ and $\theta_y = \theta \sin \phi$. Since $\overline W_{ll}^{\,\rm eikonal} - \overline W_{lk} \propto g(\vec u_{m+1},\vec u_l,\vec u_k)/(1 - \vec u_{m+1} \cdot\vec u_l)$, the main feature of $\overline W_{ll}^{\,\rm eikonal} - \overline W_{lk}$ is a singularity at $\theta_x = \theta_y = 0$. We see that the factor $g$ that multiplies the singular factor is a smooth function with a gentle peak between $\vec u_l$ and $\vec u_k$. This peak above $g=1$ represents constructive interference. When $\vec u_{m+1}$ moves outside the ``angle ordered'' region $(1 - \vec u_{m+1}\cdot \vec u_l) < (1 - \vec u_{m+1}\cdot \vec u_k)$, the factor $g$ drops below 1 and decreases to zero, representing destructive interference. We notice in Fig.~\ref{fig:softplot} that there is an enhancement of soft gluon radiation in the region between the directions of parton $l$ and parton $k$. This enhancement is known as the string effect and has been observed experimentally \cite{stringeffect}.

\section{The leading color limit}
\label{sec:color}

We have studied the spin-averaged splitting function $[\overline W_{ll} - \overline W_{lk}]$. Here $\overline W_{ll}$ describes the square of the graph for emission of a gluon from parton $l$. There are also interference graphs between emitting the gluon from parton $l$ and emitting the same gluon from parton $k$. The function $\overline W_{lk}$ describes the part of the interference graphs that we group with parton $l$. These functions give the momentum dependence. They multiply a color operator as given in Eq.~(\ref{eq:colorandspin}), 
\begin{equation}
\label{eq:color0}
-\frac{1}{2}
\left[t^{\dagger}_l \otimes t^{}_k + t^{\dagger}_k \otimes t^{}_l\right]
\;\;.
\end{equation}

We have so far not made any approximations with respect to color. Let us now take the leading color approximation. To do that, recall from Ref.~\cite{NSshower} that we use color states based on color string configurations. For instance, we could have a state $[4,5,2,3,1]$ in which 4 labels a quark, 1 labels an antiquark, and 5, 2, and 3 label gluons that connect, in that order, to a color string between the quark and antiquark. One can also have a closed string such as $(4,5,2,3,1)$ in which all of the partons are gluons. A color basis state can also consist of more than one color string connecting the partons. In general, the amplitude can have one color state $\ket{c}$ and the complex conjugate amplitude can have a color state $\bra{c'}$ with $c' \ne c$. However, in the leading color approximation we can only have $c' = c$. Additionally, in the leading color approximation we have
\begin{equation}
- t_l^\dagger \otimes t_k^{} = 
- t_k^\dagger \otimes t_l^{} =
-\frac{1}{2}
\left[t^{\dagger}_l \otimes t^{}_k + t^{\dagger}_k \otimes t^{}_l\right]
 \sim 
C_{\rm F}\, a_{lk}^\dagger \otimes a_{lk}^{}
\;\;.
\end{equation}
Here $a_{lk}^\dagger$ represents the operator that inserts gluon $m+1$ between partons $l$ and $k$ on the color string if these partons are adjacent to each other on the same color string, that is, if partons $l$ and $k$ are color connected. When $a_{lk}^\dagger$ is applied to a state $\ket{c}$ in which $l$ and $k$ are not color connected, we define $a_{lk}^\dagger \ket{c} = 0$.\footnote{We here adapt the notation of Ref.~\cite{NSshower}, where we had gluon insertion operators $a_+^\dagger(l)$ and $a_-^\dagger(l)$ that insert the gluon to the right or the left of parton $l$, respectively. If partons $l$ and $k$ are color connected, we have $a_{lk}^\dagger \ket{c} = a_+^\dagger(l) \ket{c}$ or $a_{lk}^\dagger \ket{c} = a_-^\dagger(l) \ket{c}$, depending on whether parton $k$ was to the right or left of parton $l$ along the string.} For the complex conjugate amplitude, $\bra{c}\, a_{lk}$ again gives a state with the soft gluon inserted between partons $l$ and $k$. Thus, starting with a color state $\ket{c}$ in the amplitude and $\bra{c}$ in the complex conjugate amplitude, we get zero if partons $l$ and $k$ are not color connected and we get a new color state with the soft gluon inserted between $l$ and $k$ if $l$ and $k$ are color connected. The bookkeeping on color connections is a standard part of parton shower event generators. The momentum dependent numerical factor $[\overline W_{ll} - \overline W_{lk}]$ is multiplied by a color factor $C_{\rm F}$.

This analysis has covered the case in which parton $m+1$ is a gluon, so that there are interference graphs arising from this gluon being emitted from parton $l$ in the amplitude and from parton $k$ in the complex conjugate amplitude (or the other way around). There are also graphs for which parton $m+1$ is a quark or antiquark, as described in Sec.~\ref{sec:OtherDirect}. In these cases, we have just the splitting function $\overline W_{ll}$, which multiplies the color operator $t^{\dagger}_l \otimes t^{}_l$. This operator is very simple in the leading color limit. 

Consider first the case of an initial state splitting in which $f_l = q$ and $\{\hat f_l,\hat f_{m+1}\} = \{{\rm g}, q\}$, where $q$ is a quark flavor ($\mathrm{u}$, $\bar{\rm u}$, $\mathrm{d}$, \dots). In physical time, this is a splitting ${\rm g} \to q + \bar q$, while in shower time it is a splitting $q \to {\rm g} + q$.  As discussed in Sec.~7.3 of Ref.~\cite{NSshower},
\begin{equation}
t_l^\dagger \otimes t_l 
=
C_{\rm F}\, a_{\rm g}^\dagger(l) \otimes a_{\rm g}(l)
\;\;.
\end{equation}
Here $a_{\rm g}^\dagger(l)$ represents the operator that inserts the gluon at the end of the string terminated by quark $l$ before the splitting and terminated by quark $m+1$ after the splitting.\footnote{This operator is denoted $a_{+}^\dagger(l)$ in Ref.~\cite{NSshower}.} Similarly, in the case of an initial state splitting in which $f_l = \bar q$ and $\{\hat f_l,\hat f_{m+1}\} = \{{\rm g}, \bar q\}$, we have the same result, where now $a_{\rm g}^\dagger(l)$ represents the operator that inserts the gluon at the end of the string terminated by antiquark $l$ before the splitting and terminated by antiquark $m+1$ after the splitting.

Consider next the case of an initial state splitting in which $f_l = {\rm g}$ and $\{\hat f_l,\hat f_{m+1}\} = \{\bar q, q\}$. In physical time, this is a splitting $q \to q+{\rm g}$, while in shower time it is a splitting ${\rm g} \to q + \bar q$.  As discussed in Sec.~7.3 of Ref.~\cite{NSshower}, in the leading color limit, 
\begin{equation}
\label{eq:gqqbarsplitcolor}
t^{\dagger}_l \otimes t^{}_l \sim T_{\rm R} \
a^\dagger_q(l) \otimes a_q(l)
\;\;,
\end{equation}
where $T_{\rm R} = 1/2$ and $a^\dagger_q(l)$ splits the color string at the point at which gluon $l$ attaches, creating new string ends corresponding to the quark and the antiquark. The same analysis applies for an initial state splitting with $f_l = {\rm g}$ and $\{\hat f_l,\hat f_{m+1}\} = \{q, \bar q\}$ and for a final state ${\rm g} \to q + \bar q$ splitting, for which $\{\hat f_l,\hat f_{m+1}\} = \{q, \bar q\}$.

\section{Evolution equation}
\label{sec:EvolutionEquation}

We now have the information that we need to present the formulas from Ref.~\cite{NSshower} for parton shower evolution specialized to the spin averaged, leading color approximation. In the general case, we had basis states $\sket{\{p,f,s',c',s,c\}_{m}}$ with two color configurations $\{c\}_m$ and $\{c'\}_m$, representing the color state in the amplitude and the color state in the complex conjugate amplitude, respectively, and two spin color configurations $\{s\}_m$ and $\{s'\}_m$. In this paper, we have averaged over spins, so that we can describe the evolution of the states without referring to spin at all. We also use the leading color approximation, so that we always work with states with $\{c\}_m = \{c'\}_m$. Thus our description is vastly simplified and we can work with basis states $\sket{\{p,f,c\}_{m}}$.

As in Ref.~\cite{NSshower}, we use the logarithm of the virtuality of a splitting as the evolution variable, so that a splitting of parton $l$ is assigned to a shower time $t = T_l(\{\hat p,\hat f\}_{m+1})$,
\begin{equation}
\label{eq:showertime}
T_l(\{\hat p,\hat f\}_{m+1}) = 
\log\left(\frac{Q_0^2}{|(\hat p_l +(-1)^{\delta_{l,\La} + \delta_{l,\Lb}}
\hat p_{m+1})^2 - m^2(f_l)|}\right)
\;\;,
\end{equation}
where $f_l = \hat f_l + \hat f_{m+1}$ and $Q_0^2$ is the starting virtuality scale. Shower evolution is based on the probability that, at shower time $t$, a state $\sket{\{p,f,c\}_{m}}$ that had not already split now splits to make a new state $\big|\{\hat p,\hat f,\hat c\}_{m+1}\big)$ with one more parton. This probability is represented as a matrix element of a splitting operator ${\cal H}^{(0)}_{\LI}(t)$, which is similar to the splitting operator ${\cal H}_{\LI}(t)$ of Ref.~\cite{NSshower} except that the spin averaged, leading color approximations (``(0)'') have been applied. Then ${\cal H}^{(0)}_{\LI}(t)$ operates on states $\sket{\{p,f,c\}_{m}}$ instead of the states of the full theory. We write
\begin{equation}
\label{eq:HSpinlessLeadingColor}
\begin{split}
\big(\{\hat p,\hat f,{}&\hat c\}_{m+1}\big|
{\cal H}^{(0)}_{\rm I}(t)\sket{\{p,f,c\}_{m}}
\\={}&
\sum_{l}
(m+1)\,
\frac
{n_\Lc(a) n_\Lc(b)\,\eta_{\La}\eta_{\Lb}}
{n_\Lc(\hat a) n_\Lc(\hat b)\,
 \hat \eta_{\La}\hat \eta_{\Lb}}\,
\frac{
f_{\hat a/A}(\hat \eta_{\La},\mu^{2}_{F})
f_{\hat b/B}(\hat \eta_{\Lb},\mu^{2}_{F})}
{f_{a/A}(\eta_{\La},\mu^{2}_{F})
f_{b/B}(\eta_{\Lb},\mu^{2}_{F})}
\\
&\times
\sbra{\{\hat p,\hat f\}_{m+1}}{\cal P}_{l}\sket{\{p,f\}_m}\,
\delta\!\left(
t - T_l(\{\hat p,\hat f\}_{m+1})\right)\,
\\
&\times
\biggl\{
\theta(\hat f_{m+1} = \mathrm{g})\
\sum_{\substack{k\\ k\ne l}}\
\bra{\{\hat c\}_{m+1}} a^\dagger_{lk}\ket{\{c\}_m}\,
\Phi_{lk}(\{\hat p,\hat f\}_{m+1})
\\
&\quad+
\theta(\hat f_{m+1} \ne \mathrm{g})\,\theta(\hat f_{l} = \mathrm{g})\
\bra{\{\hat c\}_{m+1}} a^\dagger_{\mathrm{g}}(l)\ket{\{c\}_m}\,
\Phi_{ll}(\{\hat p,\hat f\}_{m+1})
\\
&\quad+
\theta(\hat f_{m+1} \ne \mathrm{g})\,\theta(f_{l} = \mathrm{g})\
\bra{\{\hat c\}_{m+1}} a^\dagger_{q}(l)\ket{\{c\}_m}\,
\Phi_{ll}(\{\hat p,\hat f\}_{m+1})
\biggr\}
\;\;.
\end{split}
\end{equation}
The first line on the right hand side of this formula contains factors copied directly from Ref.~\cite{NSshower}. There is a sum over the index $l$ of the parton that splits. Then there is a ratio of parton distribution functions. This ratio is 1 for a final state splitting but different from 1 for an initial state splitting. The next line concerns the relation of the variables $\{\hat p,\hat f\}_{m+1}$ and $t$ to the variables $\{p,f\}_m$. For the flavors, this factor vanishes unless there is a QCD vertex for $f_l \to \hat f_l + \hat f_{m+1}$ and it vanishes unless $\hat f_j = f_j$ for the other partons. For an allowed relationship between $\{\hat f\}_{m+1}$ and $\{f\}_m$, the flavor factor is 1. There is a similar factor for the momenta. Given the momenta $\{p\}_m$, the momenta $\{\hat p\}_{m+1}$ must lie on a certain three dimensional surface specified by the momentum mapping ${\cal R}_l$ defined in Ref.~\cite{NSshower}. The function $\sbra{\{\hat p,\hat f\}_{m+1}}{\cal P}_{l}\sket{\{p,f\}_m}$ contains a delta function on this surface. There is also a delta function that defines the shower time $t$. Thus if we integrate $\big(\{\hat p,\hat f,\hat c\}_{m+1}\big| {\cal H}^{(0)}_{\rm I}(t) \sket{\{p,f,c\}_{m}}$ over $t$ and the momenta $\{\hat p\}_{m+1}$, we are really integrating over three variables that describe the splitting of parton $l$. 

The final factor in Eq.~(\ref{eq:HSpinlessLeadingColor}) contains three terms. Our main interest is in the first term, for $\hat f_{m+1} = {\rm g}$. There is a sum over the index $k$ of other partons in the process. These are the partons that might be connected with parton $l$ in an interference diagram. The remaining factors are rather complicated in the general case described in Ref.~\cite{NSshower}, but are quite simple in the spin averaged, leading color approximation. The factor $\bra{\{\hat c\}_{m+1}} a^\dagger_{lk}\ket{\{c\}_m}$ embodies the color considerations described in Sec.~\ref{sec:color}. It equals 1 provided two conditions hold. First, partons $l$ and $k$ must be color connected in the initial color state $\{c\}_{m}$. Second, the new color state $\{\hat c\}_{m+1}$ must be the same as $\{c\}_{m}$ with the gluon with label $m+1$ inserted between partons $l$ and $k$. If either of these conditions fails, this factor vanishes. The remaining factor is the splitting function
\begin{equation}
\Phi_{lk} \equiv C_{\rm F}\,[\overline W_{ll} - \overline W_{lk}]
\;\;.
\end{equation}
We have seen explicitly what this factor is, and have noted that $\Phi_{lk}$ is positive.

The next term in the braces in Eq.~(\ref{eq:HSpinlessLeadingColor}) applies to an initial state splitting in which $\hat f_l = {\rm g}$ and $\{f_l,\hat f_{m+1}\}$ is either $\{q,q\}$ or $\{\bar q,\bar q\}$. The color factor $\bra{\{\hat c\}_{m+1}} a^\dagger_{\mathrm{g}}(l)\ket{\{c\}_m}$ is 1 if the new color state $\{\hat c\}_{m+1}$ is the same as $\{c\}_{m}$ with the end of the string at quark or antiquark $l$ now terminated at quark or antiquark $m+1$ and the new the gluon with label $l$ inserted just next to the end of the string. Otherwise, this factor vanishes. The corresponding splitting function is
\begin{equation}
\Phi_{ll} \equiv C_{\rm F}\,\overline W_{ll}
\;\;.
\end{equation}
The final term in the braces in Eq.~(\ref{eq:HSpinlessLeadingColor}) applies to an initial state splitting in which $f_l = {\rm g}$ and $\{\hat f_l,\hat f_{m+1}\}$ is either $\{q,\bar q\}$ or $\{\bar q,q\}$. The color factor $\bra{\{\hat c\}_{m+1}} a^\dagger_{\mathrm{q}}(l)\ket{\{c\}_m}$ is 1 if the color state $\{\hat c\}_{m+1}$ is related to $\{c\}_{m}$ by cutting the color string on which parton $l$ (a gluon) lies into two strings, terminating at the new quark and antiquark. Otherwise, this factor vanishes. The corresponding splitting function is
\begin{equation}
\Phi_{ll} \equiv T_{\rm R}\,\overline W_{ll}
\;\;.
\end{equation}

We have now specified the probability that a state $\sket{\{p,f,c\}_{m}}$ splits. The probability that this state does {\em not} split between shower times $t$ and $t'$ is
\begin{equation}
\Delta^{(0)}(t,t';\{p,f,c\}_{m}) =
\exp\left(-\int_{t'}^{t} d\tau\ 
\sbra{1}{\cal H}^{(0)}_\LI(\tau)
\sket{\{p,f,c\}_m}\right)
\;\;.
\label{eq:Deltadef}
\end{equation}
Here $\sbra{1}{\cal H}^{(0)}_\LI(\tau)\sket{\{p,f,c\}_m}$ is the inclusive probability for the state $\sket{\{p,f,c\}_{m}}$ to split at time $\tau$,
\begin{equation}
\label{eq:inclusivesplit}
\sbra{1}{\cal H}^{(0)}_\LI(\tau)
\sket{\{p,f,c\}_m}
=
\frac{1}{(m+1)!}
\int \big[d\{\hat p,\hat f,\hat c\}_{m+1}\big]\
\sbra{\{\hat p,\hat f,\hat c\}_{m+1}}
{\cal H}^{(0)}_{\rm I}(t)\sket{\{p,f,c\}_{m}}
\;\;.
\end{equation}
To get the inclusive splitting probability, we have integrated over the momenta $\{\hat p\}_{m+1}$ after the splitting and summed over the flavors and colors, using the integration measure in Eq,~(3.15) of Ref.~\cite{NSshower}, supplemented by a sum over color states.\footnote{According to Eq.~(3.15) of Ref.~\cite{NSshower}, there is an extra normalization factor $\brax{\{\hat c\}_{m+1}}\ket{\{\hat c\}_{m+1}}$ in Eq.~(\ref{eq:inclusivesplit}). With our choice of the normalization of color states, this factor is not exactly 1, but it is 1 in the leading color limit.}

With these ingredients, we can describe shower evolution using the evolution equation (14.1) from Ref.~\cite{NSshower}. The evolution from a shower time $t'$ to a final time $t_{\rm f}$ at which showering is terminated is given by an operator ${\cal U}^{(0)}(t_{\rm f},t')$ that obeys\footnote{In Ref.~\cite{NSshower}, $[{\cal H}_{\LI}(\tau) - {\cal V}_{\LS}(\tau)]$ appears in place of ${\cal H}_{\LI}^{(0)}(\tau)$ here. With the leading color approximation, ${\cal V}_{\LS}(\tau) = 0$.}
\begin{equation}
\label{eq:evolutionbis}
{\cal U}^{(0)}(t_{\rm f},t') = {\cal N}^{(0)}(t_{\rm f},t')
+ \int_{t'}^{t_{\rm f}}\! d\tau\ 
{\cal U}^{(0)}(t_{\rm f},\tau)\,
{\cal H}^{(0)}_{\LI}(\tau)
\,{\cal N}^{(0)}(\tau,t')
\;\;.
\end{equation}
Here ${\cal N}^{(0)}(t',t)$ is a no-splitting operator defined by
\begin{equation}
{\cal N}^{(0)}(t',t)
\sket{\{p,f,c\}_{m}} =
\Delta^{(0)}(t',t;\{p,f,c\}_{m})
\sket{\{p,f,c\}_{m}}
\;\;.
\label{eq:Neigenvalue0}
\end{equation}
If we apply this to a state $\sket{\{p,f,c\}_{m}}$ that exists at shower time $t'$, we have
\begin{equation}
\begin{split}
\label{eq:evolutiondetail}
{\cal U}^{(0)}(t_{\rm f},t')\sket{\{p,f,c\}_{m}} ={}& 
\Delta^{(0)}(t_{\rm f},t';\{p,f,c\}_{m})\sket{\{p,f,c\}_{m}}
\\
& + 
\int_{t'}^{t_{\rm f}}\! d\tau\ 
\frac{1}{(m+1)!}
\int \big[d\{\hat p,\hat f,\hat c\}_{m+1}\big]\
{\cal U}^{(0)}(t_{\rm f},\tau)\,
\sket{\{\hat p,\hat f,\hat c\}_{m+1}}
\\&\quad \times
\sbra{\{\hat p,\hat f,\hat c\}_{m+1}}
{\cal H}^{(0)}_{\rm I}(\tau)\sket{\{p,f,c\}_{m}}
\,\Delta^{(0)}(\tau,t';\{p,f,c\}_{m})
\;\;.
\end{split}
\end{equation}
The first term gives the probability that the state does not split before shower time $t_{\rm f}$. The main evolution is represented by the second term. There is an integration over the shower time $\tau$ of the next splitting and over the splitting parameters. In an implementation of this equation, the integration would be performed by Monte Carlo integration. That is, we would choose $\tau$ and $\{\hat p,\hat f,\hat c\}_{m+1}$ with some probability density $\rho$ that 
contains delta functions that restrict $\tau$ and $\{\hat p,\hat f,\hat c\}_{m+1}$ to the allowed surface defined by the Eq.~(\ref{eq:showertime}) for $\tau$ and the momentum mapping ${\cal R}_l$. Then we multiply by a weight $w$ defined by 
\begin{equation}
\frac{1}{(m+1)!}\ {\sbra{\{\hat p,\hat f,\hat c\}_{m+1}}
{\cal H}^{(0)}_{\rm I}(\tau)\sket{\{p,f,c\}_{m}}
\,\Delta^{(0)}(\tau,t';\{p,f,c\}_{m})}
= w \times \rho
\;\;.
\end{equation}
In the present case, the integrand has two welcome features. First, it is positive. Second, using the definition of $\Delta^{(0)}$,
\begin{equation}
\begin{split}
\frac{1}{(m+1)!}&
\int_{t'}^{\infty}\! d\tau 
\int\! \big[d\{\hat p,\hat f,\hat c\}_{m+1}\big]
\sbra{\{\hat p,\hat f,\hat c\}_{m+1}}
{\cal H}^{(0)}_{\rm I}(\tau)\sket{\{p,f,c\}_{m}}
\,\Delta^{(0)}(\tau,t';\{p,f,c\}_{m})
\\ &
= 1
\;.
\end{split}
\end{equation}
Thus the function
\begin{equation}
\rho = 
\frac{\sbra{\{\hat p,\hat f,\hat c\}_{m+1}}
{\cal H}^{(0)}_{\rm I}(\tau)\sket{\{p,f,c\}_{m}}
\,\Delta^{(0)}(\tau,t';\{p,f,c\}_{m})}{(m+1)!}
\end{equation}
is positive and properly normalized to be a probability density. Using standard methods from shower Monte Carlo algorithms \cite{EarlyPythia, Gottschalk, Pythia, Herwig}, we can choose points with this probability density. Then $w=1$. With a probability $\Delta^{(0)}(t_{\rm f},t';\{p,f,c\}_{m})$, the point selected will be in the range $t_{\rm f} < \tau < \infty$. In this case, there is no splitting and we simply keep the state $\sket{\{p,f,c\}_{m}}$. This corresponds to the no splitting term in Eq.~(\ref{eq:evolutiondetail}). If $\tau < t_{\rm f}$, the state splits to $\{\hat p,\hat f,\hat c\}_{m+1}$. Then, according to Eq.~(\ref{eq:evolutiondetail}), we should apply ${\cal U}^{(0)}(t_{\rm f},\tau)$ to this state, repeating the process. Thus the evolution proceeds by what is known as a Markov chain.

The starting point for evolution is a state that is a mixture of the basis states $\sket{\{p,f,c\}_{m}}$ for $m=2$, assuming that we start with a $2 \to 2$ hard process,
\begin{equation}
\label{eq:evolutionstart}
\sket{\rho^{(0)}(0)} = \frac{1}{2!}
\int \big[d\{p,f,c\}_{2}\big]\
\sket{\{p,f,c\}_{2}}
\sbrax{\{p,f,c\}_{2}}\sket{\rho^{(0)}(0)}
\;\;.
\end{equation}
Here $\sbrax{\{p,f,c\}_{2}}\sket{\rho^{(0)}(0)}$ is obtained from the $2 \to 2$ matrix element summed over spins,\footnote{As explained in Ref.~\cite{NSshower}, we should most properly project out the component of $\ket{\ME(\{p,f\}_{2})}$ that is proportional to a color basis state $\ket{\{c\}_{2}}$ by using a dual basis state $\dualL\bra{\{c\}_{2}}$, but in the leading color limit there is no distinction between the dual basis states and the ordinary basis states.}
\begin{equation}
\label{eq:rhodef1}
\sbrax{\{p,f,c\}_{2}}\sket{\rho^{(0)}(0)} = 
\frac{f_{a/A}(\eta_{\La},\mu^{2}_{F})
f_{b/B}(\eta_{\Lb},\mu^{2}_{F})}
{4n_\Lc(a) n_\Lc(b)\,2\eta_{\La}\eta_{\Lb}p_\LA\!\cdot\!p_\LB}\,
\sum_{\{s\}_2}
\big|\brax{\{s,c\}_{2}}\ket{\ME(\{p,f\}_{2})}\big|^2
\;\;.
\end{equation}
To implement Eq.~(\ref{eq:evolutionstart}), one would choose points $\{p,f,c\}_2$ by Monte Carlo methods. This gives the starting point for the shower evolution. The state $\sket{\rho^{(0)}(0)} $ then evolves into a state
\begin{equation}
\sket{\rho^{(0)}(t_{\rm f})} = {\cal U}^{(0)}(t_{\rm f},0)\sket{\rho^{(0)}(0)}
\end{equation}
at the shower time $t_{\rm f}$ at which we choose to terminate shower evolution. At this point, as described in Ref.~\cite{NSshower}, the desired cross section is obtained by applying a hadronization model to the component states $\sket{\{p,f,c\}_{N}}$ in $\sket{\rho^{(0)}(t_{\rm f})}$, producing a hadronic state ${\cal U}^{\rm had}(\infty,t_{\rm f})\sket{\rho^{(0)}(t_{\rm f})}$. Then the desired cross section $\sigma[F_{\rm h}]$ results from applying the measurement function $F_{\rm h}$ to the hadronic states produced. Thus
\begin{equation}
\begin{split}
\label{eq:hadronization}
\sigma^{(0)}[F_{\rm h}] ={}& 
\sbra{F_{\rm h}}{\cal U}^{\rm had}(\infty,t_{\rm f})
\sket{\rho^{(0)}(t_{\rm f})}
\\
={}& 
\sum_N
\frac{1}{N!}
\int \big[d\{p,f,c\}_{N}\big]\
\sbra{F_{\rm h}}{\cal U}^{\rm had}(\infty,t_{\rm f})
\sket{\{p,f,c\}_{N}}
\sbrax{\{p,f,c\}_{N}}
\sket{\rho^{(0)}(t_{\rm f})}
\;\;.
\end{split}
\end{equation}
Just as in the parton shower evolution, the integration in Eq.~(\ref{eq:hadronization}) can be implemented by simply taking the states $\sket{\{p,f,c\}_{N}}$ generated by the shower evolution and passing them to a Monte Carlo implementation of a hadronization model. Then application of the measurement function is acheived by, for instance, putting the events into desired bins according to the momenta of the resulting hadrons.
\section{Other approaches}

In this section, we sketch the relation of the shower evolution of this paper to some other approaches to the description of parton showers. For the shake of the simplicity we work only with massless partons in this section but it is still allowed for the non-QCD particles to have non-zero masses.

\subsection{Dipole shower}

One possibility for organizing the gluon radiation in a (spin averaged, leading color) parton shower is to use the same functions that are used for organizing the subtractions in a next-to-leading order perturbative calculation. In particular, the dipole subtraction scheme of Catani and Seymour \cite{CataniSeymour} is an attractive possibility \cite{Ringberg} that has been developed as the basis for parton shower programs by Schumann and Krauss \cite{Schumann} and by Dinsdale, Ternick and Weinzierl \cite{Weinzierl}.

To see how this can work, consider the case that the emitted parton $m+1$ is a gluon, so that the splitting operator is given by the main term in Eq.~\eqref{eq:HSpinlessLeadingColor},
\begin{equation}
\label{eq:Hpair}
\begin{split}
\big(\{\hat p,\hat f,{}&\hat c\}_{m+1}\big|
{\cal H}^{(0)}_{\rm I}(t)\sket{\{p,f,c\}_{m}}
\\={}&
\sum_{l}
\sum_{\substack{k\\ k\ne l}}\
(m+1)\,
\frac
{n_\Lc(a) n_\Lc(b)\,\eta_{\La}\eta_{\Lb}}
{n_\Lc(\hat a) n_\Lc(\hat b)\,
\hat \eta_{\La}\hat \eta_{\Lb}}\,
\frac{
f_{\hat a/A}(\hat \eta_{\La},\mu^{2}_{F})
f_{\hat b/B}(\hat \eta_{\Lb},\mu^{2}_{F})}
{f_{a/A}(\eta_{\La},\mu^{2}_{F})
f_{b/B}(\eta_{\Lb},\mu^{2}_{F})}
\\
&\times
\sbra{\{\hat p,\hat f\}_{m+1}}{\cal P}_{l}\sket{\{p,f\}_m}\,
\delta\!\left(
t - T_l(\{\hat p,\hat f\}_{m+1})
\right)\,
\bra{\{\hat c\}_{m+1}} a^\dagger_{lk}\ket{\{c\}_m}
\\
&\times
\Phi_{lk}(\{\hat p,\hat f\}_{m+1})
\;\;.
\end{split}
\end{equation}
The term $l,k$ generates gluons predominately soft or collinear with parton $l$. That is because $\Phi_{lk}$ is singular when $\hat p_{m+1}$ is soft or collinear with $\hat p_l$ but finite when $\hat p_{m+1}$ is collinear with $\hat p_k$. Each term is defined with its own phase space mapping ${\cal P}_{l}$ and evolution parameter $t$. Now we can use the momentum mappings ${\cal P}^{\rm cs}_{lk}$ of Catani and Seymour. These obey  
\begin{equation}
\begin{split}
\sbra{\{\hat p,\hat f\}_{m+1}}{\cal P}^{\rm cs}_{lk}\sket{\{p,f\}_m}
\sim{}& \sbra{\{\hat p,\hat f\}_{m+1}}{\cal P}_{l}\sket{\{p,f\}_m}
\;\;\;\text{when}\;\; \hat p_{m+1} \to \lambda \hat p_{l}
\\
\sbra{\{\hat p,\hat f\}_{m+1}}{\cal P}^{\rm cs}_{lk}\sket{\{p,f\}_m}
\sim{}& \sbra{\{\hat p,\hat f\}_{m+1}}{\cal P}_{l}\sket{\{p,f\}_m} 
\\\sim{}&
\sbra{\{\hat p,\hat f\}_{m+1}}{\cal P}_{k}\sket{\{p,f\}_m}
\;\;\;\text{when}\;\; \hat p_{m+1} \to 0\;\;.
\end{split}
\end{equation}
We can also use the splitting functions $\Phi^{\rm cs}_{lk}$ of Catani and Seymour. These substitutions give
\begin{equation}
\label{eq:HpairCS}
\begin{split}
\big(\{\hat p,\hat f,{}&\hat c\}_{m+1}\big|
{\cal H}^{\rm cs}_{\rm I}(t)\sket{\{p,f,c\}_{m}}
\\={}&
\sum_{l}
\sum_{\substack{k\\ k\ne l}}\
(m+1)\,
\frac
{n_\Lc(a) n_\Lc(b)\,\eta_{\La}\eta_{\Lb}}
{n_\Lc(\hat a) n_\Lc(\hat b)\,
 \hat \eta_{\La}\hat \eta_{\Lb}}\,
\frac{
f_{\hat a/A}(\hat \eta_{\La},\mu^{2}_{F})
f_{\hat b/B}(\hat \eta_{\Lb},\mu^{2}_{F})}
{f_{a/A}(\eta_{\La},\mu^{2}_{F})
f_{b/B}(\eta_{\Lb},\mu^{2}_{F})}
\\
&\times
\sbra{\{\hat p,\hat f\}_{m+1}}{\cal P}^{\rm cs}_{lk}\sket{\{p,f\}_m}\,
\delta\!\left(
t - T_l(\{\hat p,\hat f\}_{m+1})
\right)\,
\bra{\{\hat c\}_{m+1}} a^\dagger_{lk}\ket{\{c\}_m}
\\
&\times
\Phi_{lk}^{\rm cs}(\{\hat p,\hat f\}_{m+1})
\;\;.
\end{split}
\end{equation}
The splitting operator ${\cal H}^{\rm cs}_{\rm I}(t)$ matches ${\cal H}^{(0)}_{\rm I}(t)$ in the collinear and soft limits. 

We see that the structure of shower generation using the Catani-Seymour functions is quite similar to that of this paper. It is of interest to compare the splitting functions in the soft limit, $\hat p_{m+1} \to 0$. Using the definitions in Ref.~\cite{CataniSeymour}, we have
\begin{equation}
\label{eq:softdistfromgCS}
\Phi_{lk}^{\rm cs}(\{\hat p,\hat f\}_{m+1}) \sim
\frac{4\pi\as\,C_{\rm F}\,2\hat Q^2}
{(\hat Q\!\cdot\!\hat p_{m+1})^2\ (1 - \vec u_{m+1}\!\cdot\! \vec u_l)}\
g^{\rm cs}(\vec u_{m+1},\vec u_l,\vec u_k;E_l/E_k)
\;\;,
\end{equation}
for $\hat p_{m+1} \to 0$, where
\begin{equation}
\label{eq:gcsdef}
g^{\rm cs}(\vec u_{m+1},\vec u_l,\vec u_k;E_l/E_k) =
\frac{
(1 - \vec u_{l}\!\cdot\! \vec u_k)
}
{
(E_l/E_k)(1 - \vec u_{m+1}\!\cdot\! \vec u_l)
+
(1 - \vec u_{m+1}\!\cdot\! \vec u_k)
}
\;.
\end{equation}
Here $E_l$ and $E_k$ are the energies of partons $l$ and $k$, respectively, in the rest frame of $\hat Q$, the total momentum of the final state partons. Thus $E_l/E_k = \hat p_l\cdot \hat Q/\hat p_k\cdot \hat Q$. This function is similar in form to the function $g$ of this paper, plotted in Fig.~\ref{fig:softplot}, but it depends on the ratio $E_l/E_k$. We plot it in Fig.~\ref{fig:csplots} for $E_l/E_k = 3$ and $E_l/E_k = 1/3$. We see that the Catani-Seymour functions assign little soft radiation to the more energetic of partons $l$ and $k$. More soft radiation is assigned to the less energetic parton of $l$ and $k$, with quite a lot of the radiation going in approximately the direction of the more energetic parton.
  
The the final state shower in the latest version (version 8.1) of \textsc{Pythia} \cite{SjostrandSkands,Pythia} is essentially a dipole shower as described above. In particular, the splitting function describing gluon emission in the soft limit $\hat p_{m+1} \to 0$ is that in Eq.~(\ref{eq:softdistfromgCS}) with the same function $g$ as given in Eq.~(\ref{eq:gcsdef}).

\FIGURE{
\centerline{\includegraphics[width = 8 cm]{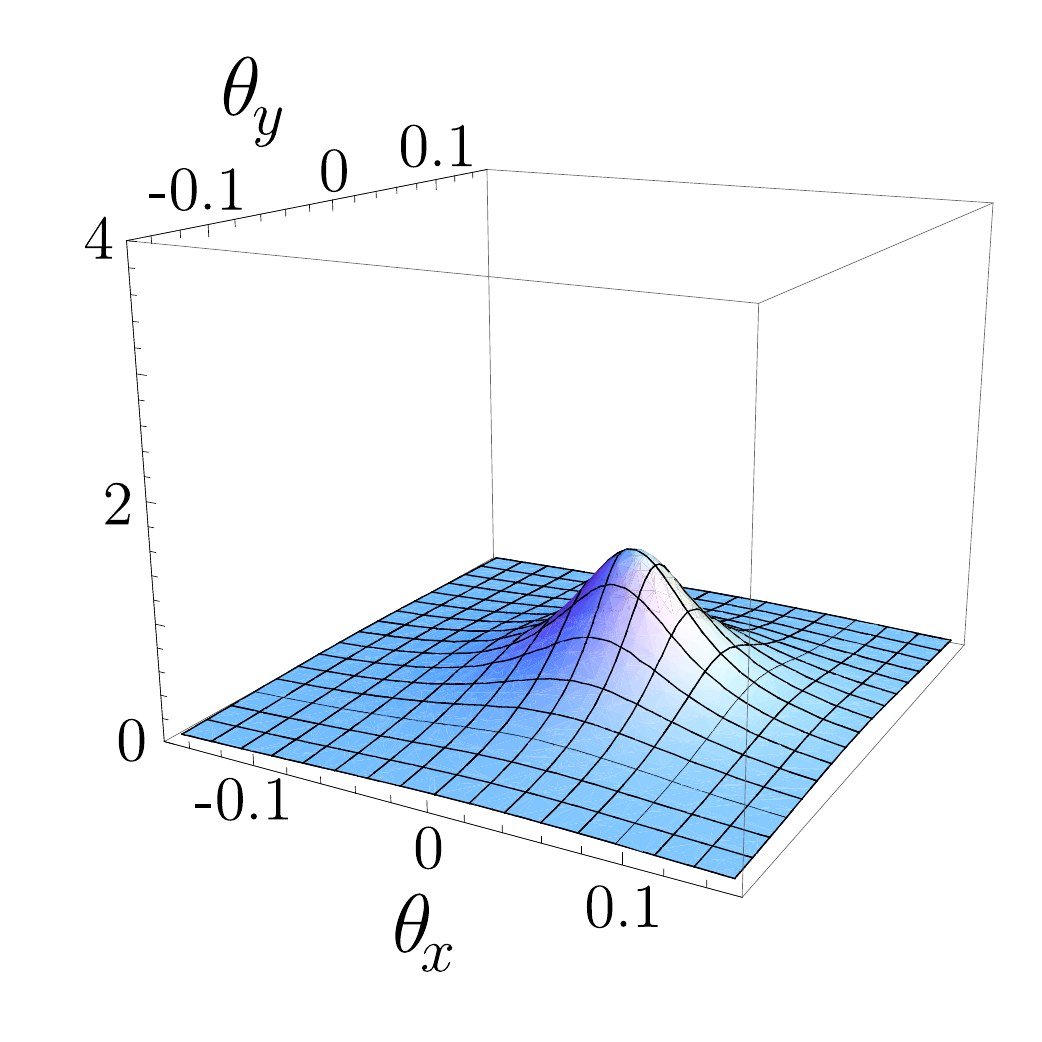}
\includegraphics[width = 7.8 cm]{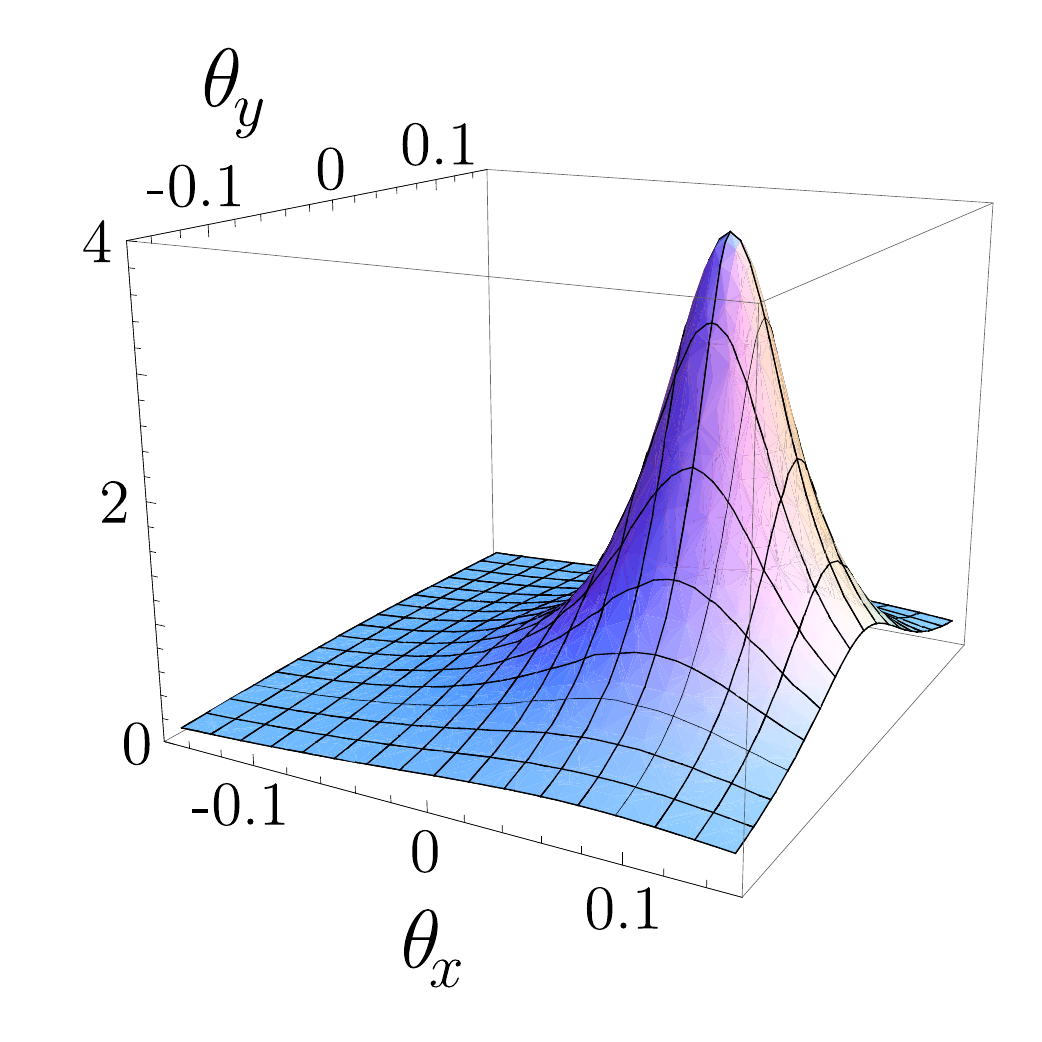}}
\caption{The soft radiation function for parton $l$ corresponding to Catani-Seymour dipole splitting. The function $g$ defined in Eq.~(\ref{eq:gdef}) and plotted in Fig.~\ref{fig:softplot} is replaced by the function $g_{\rm c.s.}$, Eq.~(\ref{eq:gcsdef}), plotted here. In the plot on the left, $E_l/E_k = 3$. In the plots on the right, $E_l/E_k = 1/3$. The plot coordinates and value of $\theta_{lk}$ are as in  Fig.~\ref{fig:softplot}.}
\label{fig:csplots}
} 

\subsection{Antenna shower}

In the method of this paper and in a dipole shower following the Catani-Seymour scheme, the creation of a new gluon is attributed to the splitting of one of the previously existing partons. This requires that for the interference graph between the amplitude for emitting the gluon from parton $l$ and the amplitude for emitting the gluon from parton $k$, one assigns a certain fraction $A$ of the graph to the splitting of parton $l$ and a fraction $1 - A$ to the splitting of parton $k$. In an antenna shower, one treats the pair of color connected partons, $l,k$ as a unit. The $l,k$ dipole constitutes an antenna that radiates the daughter gluon.\footnote{One ought to call this a dipole shower, but then one would need a new name for the kind of shower described in the previous subsection.} The pioneering development along these lines is the final state shower of \textsc{Ariadne}\cite{Ariadne}. More recent examples include those in Refs.~\cite{antenna,Vincia}. There is a corresponding subtraction scheme for next-to-leading order calculations, antenna subtraction \cite{antennasubtract}.

To define an antenna shower, we choose a momentum mapping ${\cal P}_{lk}^{\rm ant}$ with the properties previously defined and with the symmetry property
\begin{equation}
{\cal P}_{lk}^{\rm ant} = {\cal P}_{kl}^{\rm ant}
\;\;.
\end{equation}
We also redefine the shower evolution variable to be symmetric under $l \leftrightarrow k$ interchange. For instance, we could take
\begin{equation}
t = 
\log\left(\frac{Q_0^2}
{2\min[\hat p_l\!\cdot\!\hat p_{m+1},\,\hat p_k\!\cdot\!\hat p_{m+1}]}\right)
\;\;.
\end{equation}
Then we can rewrite the sum over $l$ and $k$ as a sum over pairs $l,k$, with each pair counted once, giving
\begin{equation}
\label{eq:HpairAnt}
\begin{split}
\big(\{\hat p,\hat f,{}&\hat c\}_{m+1}\big|
{\cal H}^{\rm ant}_{\rm I}(t)\sket{\{p,f,c\}_{m}}
\\={}&
\sum_{\substack{l,k\\ {\rm pairs}}}\
(m+1)\,
\frac
{n_\Lc(a) n_\Lc(b)\,\eta_{\La}\eta_{\Lb}}
{n_\Lc(\hat a) n_\Lc(\hat b)\,
 \hat \eta_{\La}\hat \eta_{\Lb}}\,
\frac{
f_{\hat a/A}(\hat \eta_{\La},\mu^{2}_{F})
f_{\hat b/B}(\hat \eta_{\Lb},\mu^{2}_{F})}
{f_{a/A}(\eta_{\La},\mu^{2}_{F})
f_{b/B}(\eta_{\Lb},\mu^{2}_{F})}
\\
&\times
\sbra{\{\hat p,\hat f\}_{m+1}}{\cal P}^{\rm ant}_{lk}\sket{\{p,f\}_m}\,
\delta\!\left(
t - \log\left(\frac{Q_0^2}
{2\min[\hat p_l\!\cdot\!\hat p_{m+1},\,\hat p_k\!\cdot\!\hat p_{m+1}]}\right)
\right)
\\
&\times
\bra{\{\hat c\}_{m+1}} a^\dagger_{lk}\ket{\{c\}_m}\
\Phi_{lk}^{\rm ant}(\{\hat p,\hat f\}_{m+1})
\;\;.
\end{split}
\end{equation}
Here $\Phi_{lk}^{\rm ant}$ can be
\begin{equation}
\Phi_{lk}^{\rm ant} = \Phi_{lk} + \Phi_{kl}
\end{equation}
or any function that matches it in the soft and collinear limits.

In the soft limit, $\hat p_{m+1} \to 0$, $\Phi_{lk}^{\rm ant}$ approaches the soft limit of the sum $\Phi_{lk} + \Phi_{kl}$, which is
\begin{equation}
\label{eq:softdistfromant}
\Phi_{lk}^{\rm ant}(\{\hat p,\hat f\}_{m+1}) \sim
\frac{4\pi\as\,C_{\rm F}\,2\hat Q^2}
{(\hat Q\!\cdot\!\hat p_{m+1})^2}\
\frac{(1 - \vec u_l\!\cdot\!\vec u_k)}
{(1 - \vec u_{m+1}\!\cdot\! \vec u_l)(1 - \vec u_{m+1}\!\cdot\! \vec u_k)}
\;\;.
\end{equation}
There is no function $g$ here. The function $g$ in the previous subsections arises from separating this into two terms, one that remains finite when $(1 - \vec u_{m+1}\!\cdot\! \vec u_k) \to 0$ and the other that remains finite when $(1 - \vec u_{m+1}\!\cdot\! \vec u_l) \to 0$.

\subsection{Angular ordering approximation}

\FIGURE{
\centerline{\includegraphics[width = 10 cm]{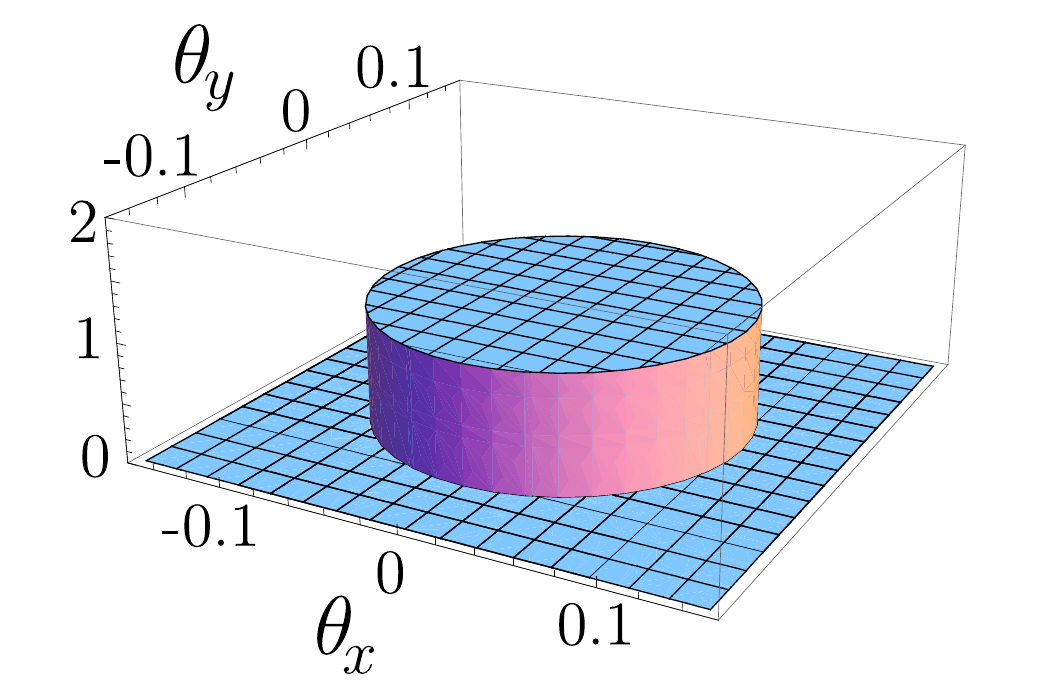}}
\caption{The angular ordering approximation. The function $g$ defined in Eq.~(\ref{eq:gdef}) and plotted in Fig.~\ref{fig:softplot} is replaced by the function $g_{\rm a.o.}$, Eq.~(\ref{eq:gaodef}), plotted here. The plot coordinates and value of $\theta_{lk}$ are as in  Fig.~\ref{fig:softplot}.}
\label{fig:angularordering}
} 

With massless kinematics, the distribution of soft radiation that is kinematically of the form for a splitting of parton $l$ is proportional to $g(\vec u_{m+1},\vec u_l,\vec u_k)/(1-\vec u_{m+1}\cdot \vec u_l)$, as given in Eq.~(\ref{eq:softdistfromg}). From the plot of $g$ in  Fig.~\ref{fig:softplot}, we see that the soft gluon radiation from partons $l$ and $k$ is approximately confined to a cone between $\vec p_l$ and $\vec p_k$. This is called ``angular ordering.'' There is also an angular ordering approximation \cite{angleorder} that is sometimes used for parton showers and, in particular, lies at the heart of \textsc{Herwig} \cite{Herwig}. With this approximation, the function $g$ in Fig.~\ref{fig:softplot} is approximated by the function plotted in Fig.~\ref{fig:angularordering},
\begin{equation}
\label{eq:gaodef}
g_{\rm a.o.}(\vec u_{m+1},\vec u_l,\vec u_k) = 
\theta(\vec u_{m+1}\cdot \vec u_l > \vec u_k\cdot \vec u_l)
\;\;.
\end{equation}
We see that in the angular region between the two hard parton directions ($\theta_x \approx 0.5, \theta_y \approx 0$ in the figures), the angular distribution of the soft radiation determined by the exact function $g$ is about twice as large as that determined by $g_{\rm a.o.}$. In other angular regions $g$ gives less soft radiation than $g_{\rm a.o.}$. The angular ordering approximation has the good feature that it gets the total amount of soft radiation right,
\begin{equation}
\label{eq:angorderintegral}
\int d\Omega_{m+1}\
\frac{g(\vec u_{m+1},\vec u_l,\vec u_k)
- g_{\rm a.o.}(\vec u_{m+1},\vec u_l,\vec u_k)}
{1 - \vec u_{m+1}\cdot \vec u_l}
= 0
\;\;.
\end{equation}
This result follows from the original construction of Refs.~\cite{angleorder}. We note, however, that the original construction involved only an integration over the azimuthal angle $\phi$, while Eq.~(\ref{eq:angorderintegral}) requires an integral over both $\theta$ and $\phi$. We have also checked Eq.~(\ref{eq:angorderintegral}) by numerical integration.

One should note that the theta function in $g_{\rm a.o.}$ restricts the emission angle of a soft gluon to be smaller than the angle between $\vec u_k$ and $\vec u_l$, where $k$ is a parton that is color connected to parton $l$. If parton $l$ is a quark, then there is only one choice for $k$. However, if parton $l$ is a gluon, then there are two color connected partons. Then there are two contributions with separate angle restrictions.

\section{Conclusions}
\label{sec:conclusions}

In Ref.~\cite{NSshower}, we presented evolution equations that represent a leading order parton shower including quantum interference, spin, and color. We did not, however, present a way to implement the integrations implied by these equations in a fashion that would be practical for more than a few partons. The idea behind the evolution equations was to make just one approximation: that the virtualities in successive splittings are strongly ordered. 

Typical Monte Carlo event generators, such as \textsc{Pythia} \cite{Pythia}, \textsc{Ariadne} \cite{Ariadne}, \textsc{Herwig} \cite{Herwig}, and \textsc{Sherpa} \cite{Sherpa}, make additional approximations. In particular, they typically average over parton spins and take the leading term in an expansion in $1/N_\Lc^2$, where $N_\Lc = 3$ is the number of colors. Our aim in this paper has been to work out how the general formalism could work as a practical calculation if we make the further approximations of averaging over parton spins\footnote{More precisely, we average over the spins of a parton before it splits and sum over the spins of daughter partons.} and of keeping only the leading order in $1/N_\Lc^2$. We do, however, keep some aspects of quantum interference in that the interference graphs between the emission of a soft gluon from parton $l$ and the emission of the soft gluon from another parton $k$ are accounted for.

The result is an algorithm that is similar to what is done in widely used parton shower event generators in that the calculation can be implemented as a Markov chain, as described in Sec.~\ref{sec:EvolutionEquation}. The form of the evolution is perhaps most similar to that in the dipole showers of Refs.~\cite {Schumann} and \cite{Weinzierl} and is also similar to the $k_\perp$ version of \textsc{Pythia} \cite{SjostrandSkands}. One can think of the basic object that splits as not one parton, but two partons, $l$ and $k$, that are next to each other along a color string. This basic object is often referred to as a color dipole. When we incorporate the joint splitting of partons $l$ and $k$, there is a contribution to the splitting probability that corresponds to the square of the amplitude for parton $l$ to split. There is another contribution to the splitting probability that corresponds to the square of the amplitude for parton $k$ to split. Then there are two contributions that correspond to the interference of these amplitudes. We reorganize the four terms into two terms. One is kinematically of the form for a splitting of parton $l$, while the other is kinematically of the form for a splitting of parton $k$. This is rather similar to the structure of the dipole subtraction scheme for next-to-leading order calculations proposed by Catani and Seymour \cite{CataniSeymour}, which has been implemented for parton showers in two recent papers \cite{Schumann,Weinzierl}.  

There are differences between the shower formulation used here and that in, say, the dipole showers of Refs.~\cite {Schumann} and \cite{Weinzierl}. The splitting functions are different. In particular, we have separate formulations for the interference graphs (based on the simple eikonal approximation) and for the direct graphs, for which our splitting functions are quite directly read off from the Feynman graphs with a minimal approximation applied where an off-shell mother parton attaches to a hard scattering amplitude. The momentum mapping functions, which were presented in Ref.~\cite{NSshower}, are also different. They are similar to the Catani-Seymour momentum mappings in that they are systematically defined, invertible mappings, but they have the advantage that the form of the mapping depends on the parton index $l$ but not on the index $k$ of the partner parton.

We have seen that the leading color, spin averaged shower of this paper has a structure similar to that implemented in standard parton shower event generators. In particular, this simple shower can be implemented using a Markov chain. The full shower formalism of Ref.~\cite{NSshower} is more general than the simple shower in that parton spin and color correlations are included. We anticipate that the full formalism will be more difficult than the simple version to implement in a practical fashion. However, we anticipate that one can use the simple shower as a basis for a systematically improvable approximation to the full shower. The idea would be to start with the simple shower and provide parameters that remove the approximations gradually, so that the result is still approximate but the approximation is systematically improvable as computer resources allow. We expect to return to this subject in future papers.


\acknowledgments{
This work was supported in part the United States Department of Energy and by the Hungarian Scientific Research Fund grant OTKA T-60432.
}

\appendix

\section{The remaining splitting functions}
\label{app:OtherSplittings}

In this section we record the spin averaged splitting functions $\overline W_{ll}$ for the cases in which $\hat f_{m+1}\ne {\rm g}$, which were not covered in the main body of the paper. We use the general definition (\ref{eq:barwlldef}) of $\overline W_{ll}$ together with the formulas from Ref.~\cite{NSshower} for the splitting amplitudes $v_l$.

We first consider a final state splitting with $\{f_l,\hat f_l,\hat f_{m+1}\} = \{{\rm g},q,\bar q\}$ where $q$ is a quark flavor and $\bar q$ is the corresponding antiflavor. A straightforward calculation gives
\begin{equation}
\overline W_{ll}(\{\hat f,\hat p\}_{m+1}) = 
\frac{8\pi\as}{(\hat p_{l}+\hat p_{m+1})^{2}}
\left(1+ 
\frac{2\ \hat p_{l}\!\cdot\!D(p_{l},\hat Q)\!\cdot\!\hat p_{m+1}}
{(\hat p_{l}+\hat p_{m+1})^{2}}
\right)\;\;.
\end{equation}
For an initial state splitting with $\{f_l,\hat f_l,\hat f_{m+1}\} = \{{\rm g},q,\bar q\}$, we find
\begin{equation}
\overline W_{ll}(\{\hat f,\hat p\}_{m+1}) = 
\frac{8\pi\as}{(\hat p_{l}-\hat p_{m+1})^{2}}
\left(-1+ 
\left(\frac{\hat p_{l}\!\cdot\!n_{l}}
{(\hat p_{l}-\hat p_{m+1})\!\cdot\!n_{l}}\right)^{2}
\frac{2\ \hat p_{m+1}\!\cdot\!D(p_{l},\hat Q)\!\cdot\!\hat p_{m+1}}
{(\hat p_{l}-\hat p_{m+1})^{2}}
\right)\;\;.
\end{equation}
Here $n_l = p_\LB$ for $l = \La$ and $n_l = p_\LA$ for $l = \Lb$. The same result holds for an initial state splitting with $\{f_l,\hat f_l,\hat f_{m+1}\} = \{{\rm g},\bar q,q\}$.

We consider next an initial state splitting with $\{f_l,\hat f_l,\hat f_{m+1}\} = \{q, {\rm g},q\}$. A straightforward calculation gives
\begin{equation}
\overline w_{ll}(\{\hat f,\hat p\}_{m+1}) = 
\frac{4\pi\as}{\hat p_{l}\!\cdot\!\hat p_{m+1}}
\left(
\frac{\hat p_{l}\!\cdot\!n_{l}}{p_{l}\!\cdot\!n_{l}}
- \frac{(\hat p_{l} - \hat p_{m+1})\!\cdot\!n_{l}}{p_{l}\!\cdot\!n_{l}}\,
\frac{\hat p_{m+1}\!\cdot\!D(\hat p_{l}, \hat Q)\!\cdot\!\hat p_{m+1}}
{\hat p_{l}\!\cdot\!\hat p_{m+1}}
\right)\;\;.
\end{equation}
Again, $n_\La = p_\LB$ and $n_\Lb = p_\LA$. The same result holds for an initial state splitting with $\{f_l,\hat f_l,\hat f_{m+1}\} = \{\bar q, {\rm g},\bar q\}$.

This completes the analysis of $\overline W_{ll}$ for cases in which $\hat f_{m+1}\ne {\rm g}$.

\end{document}